\documentclass{paper}

\renewcommand{\d}{\mathrm{d}}

\begin{document}

\title{Numerical Methods in Gravitational Lensing}
\author{Matthias Bartelmann\\
  MPI f\"ur Astrophysik, P.O.~Box 1317, D--85740 Garching, Germany}

\date{\em Proceedings Contribution, Gravitational Lensing Winter
  School, Aussois 2003}

\begin{abstract}
Most problems in gravitational lensing require numerical
solutions. The most frequent types of problems are (1) finding
multiple images of a single source and classifying the images
according to their properties like magnification or distortion; (2)
propagating light rays through large cosmological simulations; and (3)
reconstructing mass distributions from their tidal field. This lecture
describes methods for solving such problems. Emphasis is put on using
adaptive-grid methods for finding images, issues of spatial resolution
and reliability of statistics for weak lensing by large-scale
structures, and methodical questions related to shear-inversion
techniques.
\end{abstract}

\maketitle

\section{Introduction}

Only for very special lens models can numerical methods be avoided in
gravitational lensing studies. There are three essential reasons for
that. One is the non-linearity of gravitational lensing, i.e.~the fact
that image and source positions are related to one another in a
non-linear fashion. This gives rise to the well-known phenomena of
mutiple imaging, strong image distortions, and so forth. The second
reason is that lenses exist which are themselves best described by
numerical models. Galaxy clusters are one example, lensing by
large-scale structures is another. Although it is true that many
aspects of gravitational lensing by large-scale structures can be
derived analytically, detailed simulations require numerical
techniques. The third reason is that the interpretation of
gravitational lensing effects or events often require the application
of sophisticated algorithms to ever growing amounts of data. One
example is the reconstruction of the projected mass density
distribution of a galaxy cluster from the observed image distortions
due to gravitational shear.

Needless to say, there are many more aspects of numerical methods
related to gravitational lensing than I can cover in this review. An
outstanding example are the highly elaborate methods that have been
developed over recent years for determining image shapes of faint
background galaxies on CCD frames, and for extracting the
gravitational shear signal from them. This is a whole branch of data
analysis on its own. Here, I can only deal with numerical methods for
relating mass distributions to their gravitational lensing effects.

Consequently, the outline of this lecture is as follows: First, I
shall discuss methods for studying individual lenses, i.e.~their
imaging properties, their critical curves and caustics. In particular,
the use of adaptive grids and techniques for searching and
characterising images will be discussed. Second, I shall describe how
extended lenses can be treated numerically using the multiple-lens
plane theory. This will lead to the basic equations for tracing light
rays through (simulated) cosmological volumes. A large fraction of the
discussion will be devoted to issues of resolution and noise, and to
spurious effects in simulated lensing statistics. Finally, third, I
shall describe inversion techniques, i.e.~methods for reconstructing
the projected mass distribution of lenses whose distortion has been
measured. The classic Kaiser-Squires method will be described, and
also maximum-likelihood techniques and maximum-entropy methods.

General lensing theory and the theory of weak lensing are covered by
Koenraad Kuijken's and Peter Schneider's lectures in this
volume. Basic references on lensing include the textbook by Schneider
et al.~(1992) and the lecture by Narayan \& Bartelmann (1999), reviews
of weak lensing are Mellier (1999) and Bartelmann \& Schneider (2001).

\section{Individual Lenses}
\subsection{Assumptions}

A brief reminder of the basic assumptions underlying the theory of
individual lenses may be in order. There are three main
assumptions. First, the Newtonian gravitational potential of the lens
be small, $|\Phi|\ll c^2$. Second, velocities in the gravitational
lens system, both of constituents within the lenses and of the lenses
with respect to the rest frame of the microwave background, be small
$v\ll c$. Third, the extent of the lenses along the line-of-sight be
small compared to the other distances in the system, which are usually
cosmological and thus comparable to the Hubble radius,
$c/H_0=3\,h^{-1}\,\mathrm{Gpc}$, with $H_0$ being the Hubble constant
and $h=H_0/100\,\mathrm{km\,s^{-1}\,Mpc^{-1}}$.

It is worth noting how well these assumptions are satisfied in
ordinary lensing situations. Consider a galaxy cluster with mass
$M=10^{15}\,h^{-1}M_\odot$. Assuming spherical symmetry, the Newtonian
potential at a distance $R=1\,h^{-1}\,\mathrm{Mpc}$ from its centre
is
\begin{equation}
  |\Phi|\approx\frac{G\,M}{R}\approx
  (2\times10^3\,\mathrm{km\,s^{-1}})^2\;,
\label{eq:1}
\end{equation}
evidently much smaller than the speed of light squared. A typical
length scale for the radius of a galaxy cluster is
$1-1.5\,h^{-1}\,\mathrm{Mpc}$, which is several hundred times smaller
than typical distances in a cluster-lensing system. Finally, peculiar
velocities of galaxy clusters with respect to the Hubble flow are of
order several hundred $\mathrm{km\,s^{-1}}$, and typical velocities
of galaxies within galaxy clusters reach of order
$10^3\,\mathrm{km\,s^{-1}}$, but both velocities are way below the
speed of light. The above assumptions hold even better for lensing by
galaxies, of course.

\begin{figure}[ht]
  \centerline{\includegraphics[width=\hsize]{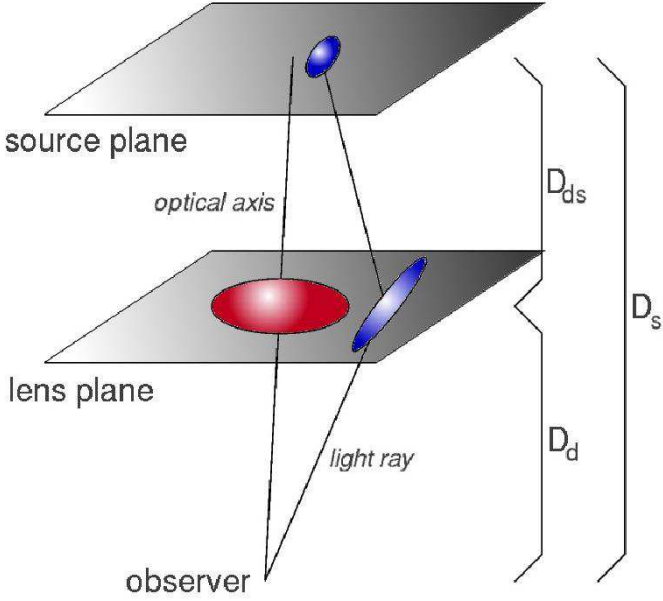}}
\caption{Schematic view of a gravitational lens system. The lens is
  projected onto the lens plane perpendicular to the line-of-sight,
  sources are located on the parallel source plane. There are three
  distances required to describe the geometry of the system, i.e.~the
  distances $D_\mathrm{d,s,ds}$ between the observer and the lens, the
  observer and the source, and between the lens and the source,
  respectively. Due to space-time curvature, these distances are
  generally not additive.}
\label{fig:1}
\end{figure}

We can thus safely assume the above conditions to be satisfied. It is
then possible to project the lensing mass distribution onto a plane
perpendicular to the line-of-sight, the lens plane, and describe it by
its surface mass density $\Sigma$. Sources are assumed to be located
on a corresponding plane, the source plane. A typical lens system is
sketched in Fig.~\ref{fig:1}.

The three distances $D_\mathrm{d,s,ds}$ shown in Fig.~\ref{fig:1} and
explained in its caption are generally not additive because of
space-time curvature, thus $D_\mathrm{s}\ne
D_\mathrm{d}+D_\mathrm{ds}$ in contrast to flat space-time.

\subsection{Coordinates and Notation}

Let us now introduce physical coordinates $\vec\xi$ and $\vec\eta$ on
the lens and source planes, respectively. Alternatively, it is often
convenient to introduce angular coordinates $\vec\theta$ and
$\vec\beta$, which are obviously related to $\vec\xi$ and $\vec\eta$
through
\begin{equation}
  \vec\xi=D_\mathrm{d}\,\vec\theta\;,\quad
  \vec\eta=D_\mathrm{s}\,\vec\beta\;.
\label{eq:2}
\end{equation}
Dimensional coordinates are of course not suitable for numerical
calculations, which can only handle numbers. We thus have to introduce
a length scale $\xi_0$, or alternatively an angular scale $\theta_0$,
in the lens plane. This length scale is so far \emph{arbitrary}. It
implies a length or angular scale
\begin{equation}
  \eta_0=\frac{D_\mathrm{s}}{D_\mathrm{d}}\,\xi_0\;,
  \quad\mbox{or}\quad
  \beta_0=\frac{\eta_0}{D_\mathrm{s}}=
  \frac{\xi_0}{D_\mathrm{d}}=\theta_0
\label{eq:3}
\end{equation}
in the source plane. Dimension-less coordinates are then defined by
\begin{equation}
  \vec x=\frac{\vec\xi}{\xi_0}=\frac{\vec\theta}{\theta_0}\;,
  \quad\mbox{or}\quad
  \vec y=\frac{\vec\eta}{\eta_0}=\frac{\vec\beta}{\theta_0}
\label{eq:4}
\end{equation}
in the lens and source planes, respectively. The numerical code will
have to deal with the dimension-less vectors $\vec x$ and $\vec y$. It
helps numerical accuracy greatly if these numbers are of order
unity. Thus, the first challenge in setting up a lensing simulation is
to choose an appropriate length- or angular scale $\xi_0$ or
$\theta_0$, which should both be adapted to the physical problem at
hand, and to the requirement that numerical codes work most accurately
if the numbers they are dealing with are neither too large nor too
small, compared to machine accuracy. Choosing unappropriate length
scales can, for instance, render image searches unsuccessful.

\subsection{The Lensing Potential}

It will be convenient for the following discussion to introduce the
lensing potential $\psi$ as the basic physical quantity for lensing
studies. It is the scaled, projected Newtonian gravitational potential
of the lens,
\begin{equation}
  \psi(\vec x)=\frac{2}{c^2}\,
  \frac{D_\mathrm{d}D_\mathrm{ds}}{\xi_0^2\,D_\mathrm{s}}\,
  \int\,\Phi(\xi_0\vec x,l)\,\d l\;.
\label{eq:5}
\end{equation}
The so-called \emph{reduced} (i.e.~appropriately scaled) deflection
angle is the gradient of the potential,
\begin{equation}
  \vec\alpha(\vec x)=\nabla_{\vec x}\psi(\vec x)\;,
\label{eq:6}
\end{equation}
and the lensing convergence (i.e.~the appropriately scaled
surface-mass density) is
\begin{equation}
  \kappa(\vec x)=\frac{1}{2}\,\nabla_{\vec x}^2\psi(\vec x)=
  \frac{1}{2}\,\nabla_{\vec x}\cdot\vec\alpha(\vec x)\;.
\label{eq:7}
\end{equation}
Finally, the gravitational tidal field is described by the
two-component shear,
\begin{equation}
  \gamma_1(\vec x)=\frac{1}{2}\left(\psi_{,11}-\psi_{,22}\right)=
  \frac{1}{2}\left(\alpha_{1,1}-\alpha_{2,2}\right)\;,\quad
  \gamma_2(\vec x)=\psi_{,12}=\alpha_{1,2}\;,
\label{eq:8}
\end{equation}
where the convention was used that $f_{i,j}$ is the derivative of the
$i$-th component of $\vec f$ with respect to the coordinate $x_j$. It
is important to note that the fact that all lensing quantities can be
derived from the scalar lensing potential establishes relations
between all of them. This will be exploited several times later.

Note that the lensing quantities must be rescaled in case the
coordinate scale $\xi_0$ is changed. Suppose $\xi'_0$ is introduced
instead of $\xi_0$. Since the physical surface-mass density of the
lens must remain the same at any given physical location, the
reduced deflection angle must transform as
\begin{equation}
  \vec\alpha(\vec x')=\frac{\xi_0'}{\xi_0}\,\vec\alpha(\vec x)\;,
\label{eq:9}
\end{equation}
and convergence and shear transform as
\begin{equation}
  \left[\kappa(\vec x')\,,\,\gamma_i(\vec x')\right]=
  \left(\frac{\xi_0'}{\xi_0}\right)^2\,
  \left[\kappa(\vec x)\,,\,\gamma_i(\vec x)\right]\;.
\label{eq:10}
\end{equation}

\subsection{Imaging}

Suppose now we were given some description of the lensing potential
$\psi(\vec x)$, or of the deflection angle $\vec\alpha(\vec x)$. This
description could be an analytical formula, or it could be in form of
an array, i.e.~a set of numbers given at grid points $(x_i,x_j)$. We
wish to know how the given lens images its background.

\begin{figure}[ht]
  \includegraphics[width=0.49\hsize]{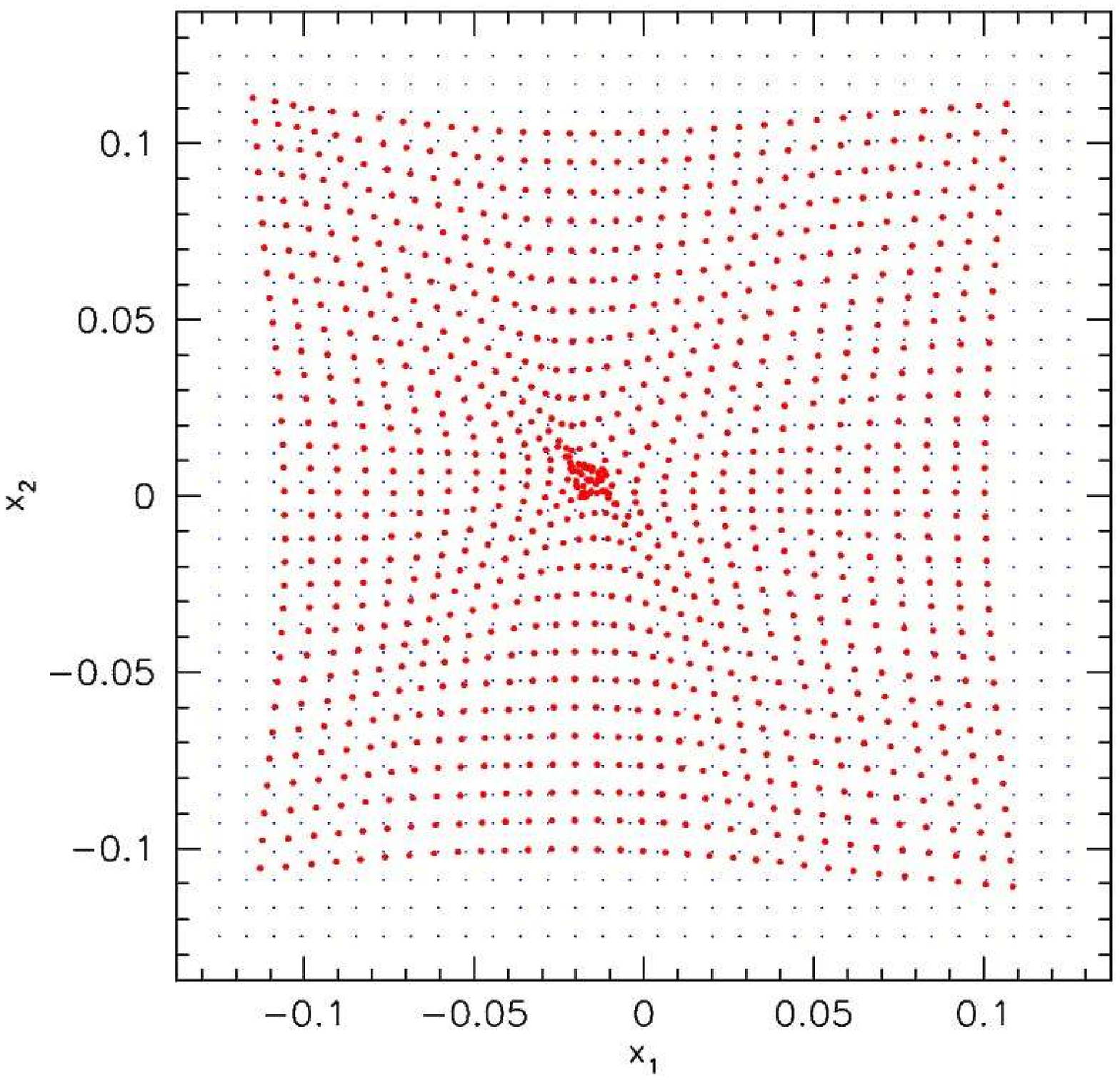}\hfill
  \includegraphics[width=0.49\hsize]{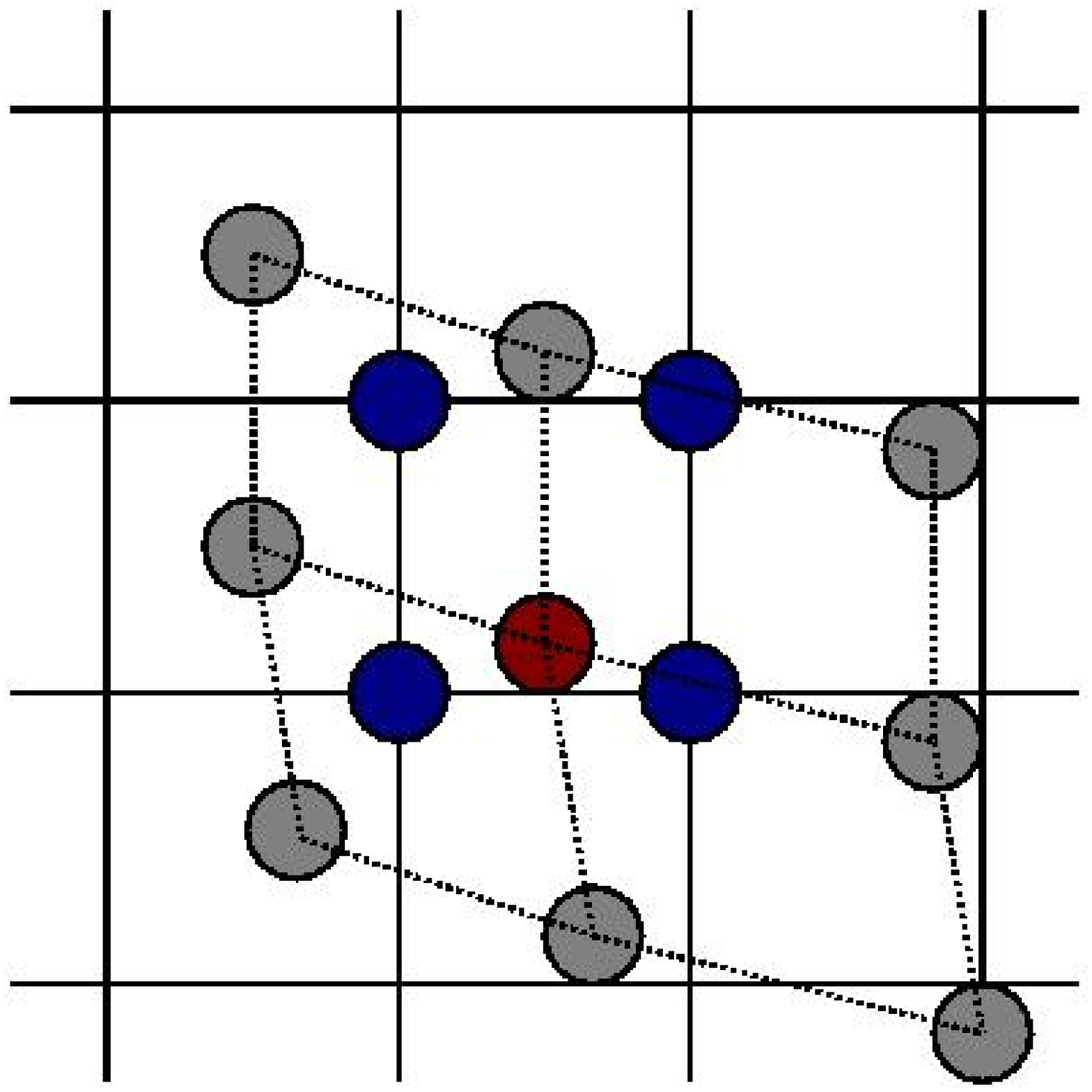}
\caption{\emph{Left panel:} A regular grid in the lens plane (blue
  dots) is mapped onto the source plane (red dots) using a numerical
  description of a deflection-angle field. Distortions are clearly
  visible. \emph{Right panel:} For each point in the lens plane, those
  points of a regular grid in the source plane (blue) are searched
  which surround its mapped point in the source plane (red).}
\label{fig:2}
\end{figure}

We introduce a coordinate grid $\vec x_{ij}$ on the lens plane subject
to the condition that it be sufficiently well resolved. This means
that the smallest features in the lens must be covered by at least a
few grid points. Since we are given the deflection angle as a function
of position, we can compute a deflection-angle grid,
$\vec\alpha_{ij}=\vec\alpha(\vec x_{ij})$. The mapped grid on the
source plane is then simply $\vec y_{ij}=\vec
x_{ij}-\vec\alpha_{ij}$. This mapped grid will appear as a distorted
image of the regular grid in the lens plane, as the example in the
left panel of Fig.~\ref{fig:2} shows.

\begin{figure*}[ht]
  \includegraphics[width=\hsize]{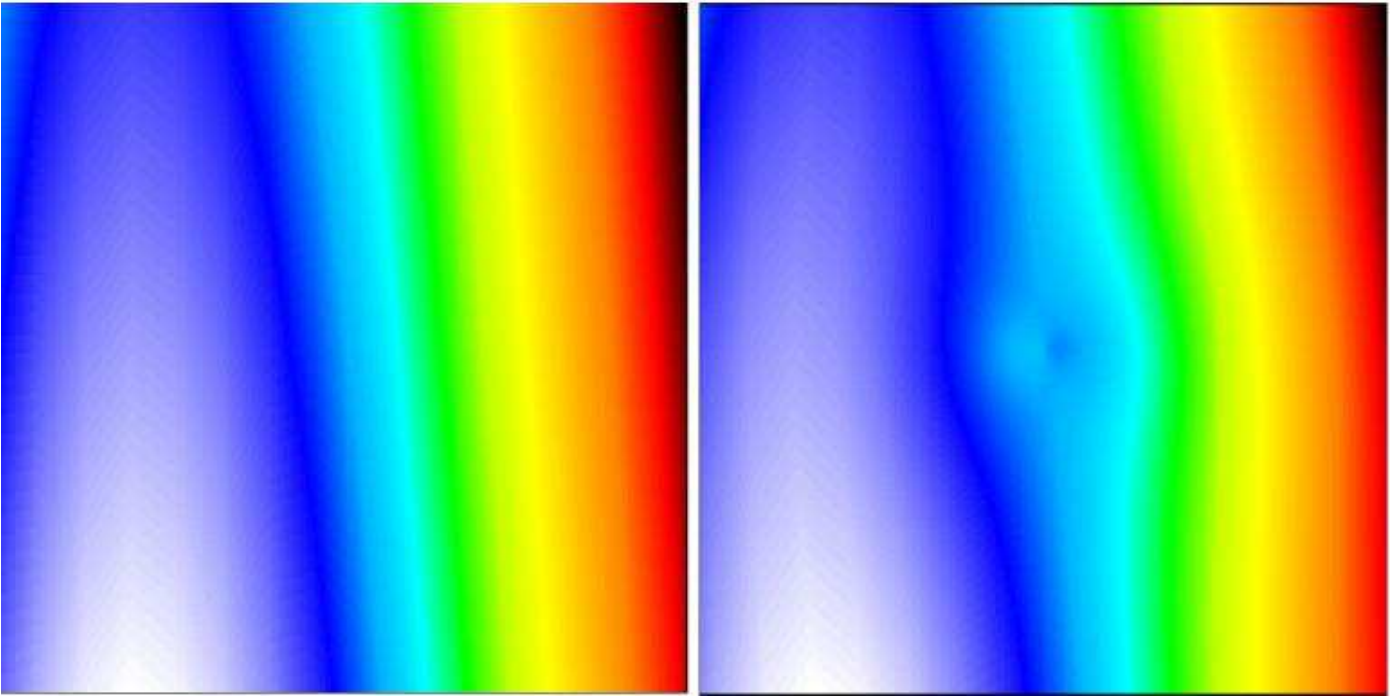}
\caption{\emph{Left panel:} A simulated CMB temperature fluctuation
  field of $10'\times10'$ size. \emph{Right panel:} The same field,
  lensed by a galaxy cluster, which imprints a characteristic pattern
  on the temperature fluctuations.}
\label{fig:3}
\end{figure*}

The mapping process must now be reversed in order to obtain an image
created by the lens. For doing so, the \emph{source} plane is first
covered with a regular grid, $\vec y_{ij}'$. Next, we loop over all
grid points $\vec x_{ij}$ in the \emph{lens} plane and find its mapped
source point $\vec y_{ij}$ in the source plane, and search for the
nearest neightbours $\vec y_{kl}'$ surrounding $\vec y_{ij}$ in the
source plane. This is illustrated in the right panel of
Fig.~\ref{fig:2}. The surface brightness of the source, known at the
positions $\vec y_{kl}'$, can then be interpolated to $\vec y_{ij}$
and the result assigned to the image point $\vec x_{ij}$. That way,
the surface brightness at all points in the lens plane can be
determined, and thus the lensed image be constructed. Fig.~\ref{fig:3}
shows an example.

The left panel of the figure shows a simulated CMB temperature
fluctuation field of $10'\times10'$ angular size. The temperature
increases from white to red. In essence, the temperature fluctuation
corresponds to a fairly smooth gradient across the field. The right
panel shows the gravitational lensing signature imprinted on the CMB
at such angular scales by a galaxy cluster. The temperature visible at
an angular position $\vec\theta$ on the sky, $T'(\vec\theta)$, is
related to the intrinsic temperature $T$ through
$T'(\vec\theta)=T[\vec\theta-\vec\alpha(\vec\theta)]$. Thus, the light
deflection by the cluster causes the visible temperature distribution
to be rearranged, yielding a highly specific pattern (Seljak \&
Zaldarriaga 2000).

\subsection{Critical Curves and Caustics}

As mentioned in the introduction, the deflection-angle field contains
full information on the lensing mass distribution. All other
quantities like convergence and shear, but also image magnifications,
follow from the deflection angle via differentiation. It is thus a
common task to compute numerical derivatives.

\begin{figure}[ht]
  \includegraphics[width=0.49\hsize]{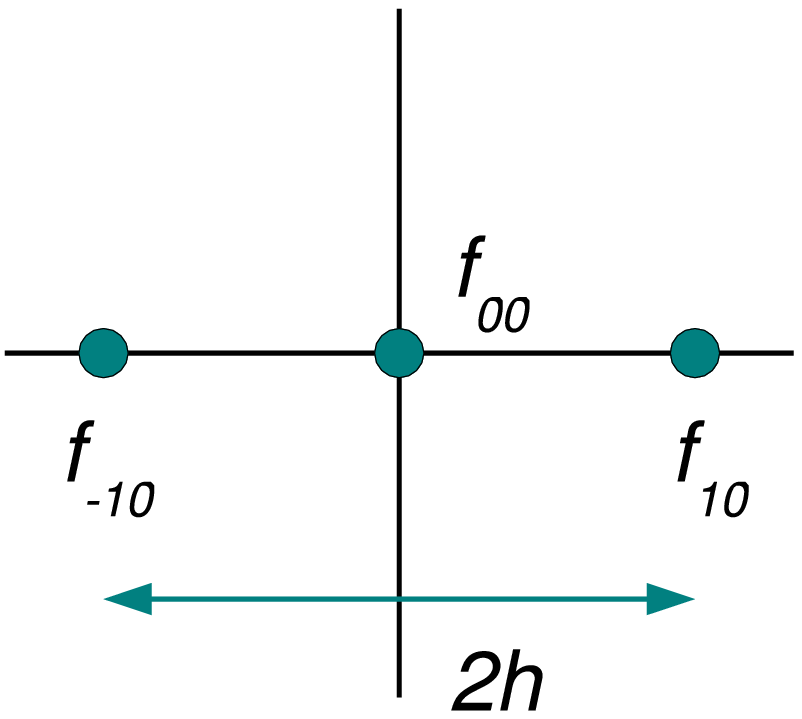}\hfill
  \includegraphics[width=0.49\hsize]{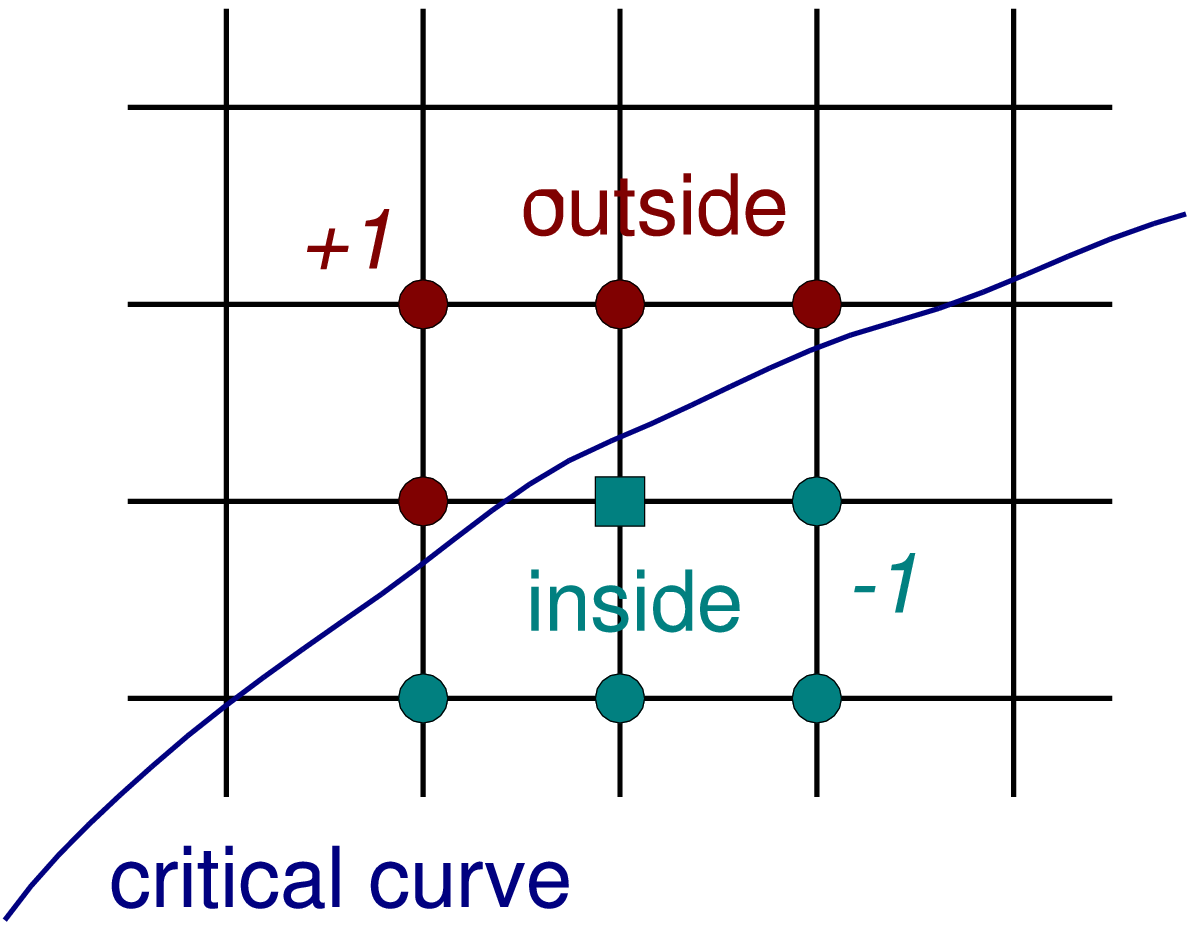}
\caption{\emph{Left panel:} Second-order numerical differentiation
  using centred differences. \emph{Right panel:} A simple method for
  finding points in the lens plane next to a critical curve uses sign
  changes of the Jacobian determinant between the point considered and
  its four nearest neighbours.}
\label{fig:4}
\end{figure}

Suppose a function $f(\vec x)$ is tabulated on a grid, so that we are
given the values $f_{ij}$ at the grid points $\vec x_{ij}$. The
derivative of $f(\vec x)$ at a particular point $\vec x_{00}$ in the
first coordinate direction is approximated by
\begin{equation}
  \left.\frac{\partial f(\vec x)}{\partial x_1}
  \right|_{\vec x_{00}}=\frac{1}{2h}\,\left(f_{10}-f_{-10}\right)+
  \mathcal{O}(h^2)\;\,
\label{eq:11}
\end{equation}
where $h$ is the separation of the grid points in the chosen
direction; cf.~the left panel of Fig.~\ref{fig:4}. This centred
difference has the advantage compared to the more straightforward
one-sided differences $f_{10}-f_{00}$ or $f_{00}-f_{-10}$ of being
second-order in the grid separation $h$. There are higher-order
differencing schemes using function values at more than two grid
points, but the second-order scheme is usually sufficient. No lensing
quantity should vary strongly between two adjacent grid points because
otherwise the resolution of the grid would be grossly insufficient.

We will typically need derivatives of the deflection angle
$\vec\alpha$. Since $\vec\alpha$ is itself the gradient of a scalar
potential, its derivatives must satisfy
$\alpha_{1,2}=\psi_{,12}=\psi_{,21}=\alpha_{2,1}$. It is thus usually
preferable to check that this relation is satisfied within numerical
accuracy, and to use $(\alpha_{1,2}+\alpha_{2,1})/2$ instead of either
$\alpha_{1,2}$ or $\alpha_{2,1}$ alone.

Critical curves in the lens plane are defined by the condition that
the Jacobian determinant of the lens mapping vanish there,
$\det\mathcal{A}(\vec x)=0$. The elements of the Jacobian matrix are
$\mathcal{A}_{ij}=\delta_{ij}-\alpha_{i,j}$, thus the Jacobian
determinant is
\begin{equation}
  D\equiv\det\mathcal{A}=
  (1-\alpha_{1,1})(1-\alpha_{2,2})-\alpha_{1,2}^2\;.
\label{eq:12}
\end{equation}
It can be computed once the (numerical) derivatives of the both
deflection-angle components have been determined.

One method of identifying grid points in the lens plane next to the
critical curve proceeds as follows. Let $S=\mathrm{sign}(D)$, and
consider one particular grid point $\vec x_{00}$ in the lens
plane. The point is next to the critical curve if, and only if, the
sign of the Jacobian determinant changes between it and one or more of
its nearest neighbours. Hence, if the condition
\begin{equation}
  S_{00}(S_{-10}+S_{10}+S_{0-1}+S_{01})<4
\label{eq:13}
\end{equation}
is satisfied, the grid point $\vec x_{00}$ is next to a critical curve
(cf.~the right panel of Fig.~\ref{fig:4}). Of course, $\vec x_{00}$ is
not itself \emph{on} the critical curve, but to the positional
accuracy determined by the grid resolution, the position of the
critical curve can be constrained that way. Points on the source plane
next to the caustic curve are then easily found via the lens equation,
$\vec y_{\mathrm{C}ij}=\vec x_{\mathrm{C}ij}-\vec\alpha(\vec
x_{\mathrm{C}ij})$, where the $\vec x_{\mathrm{C}ij}$ are the grid
points in the lens plane next to critical curves.

\begin{figure*}[ht]
  \includegraphics[width=\hsize]{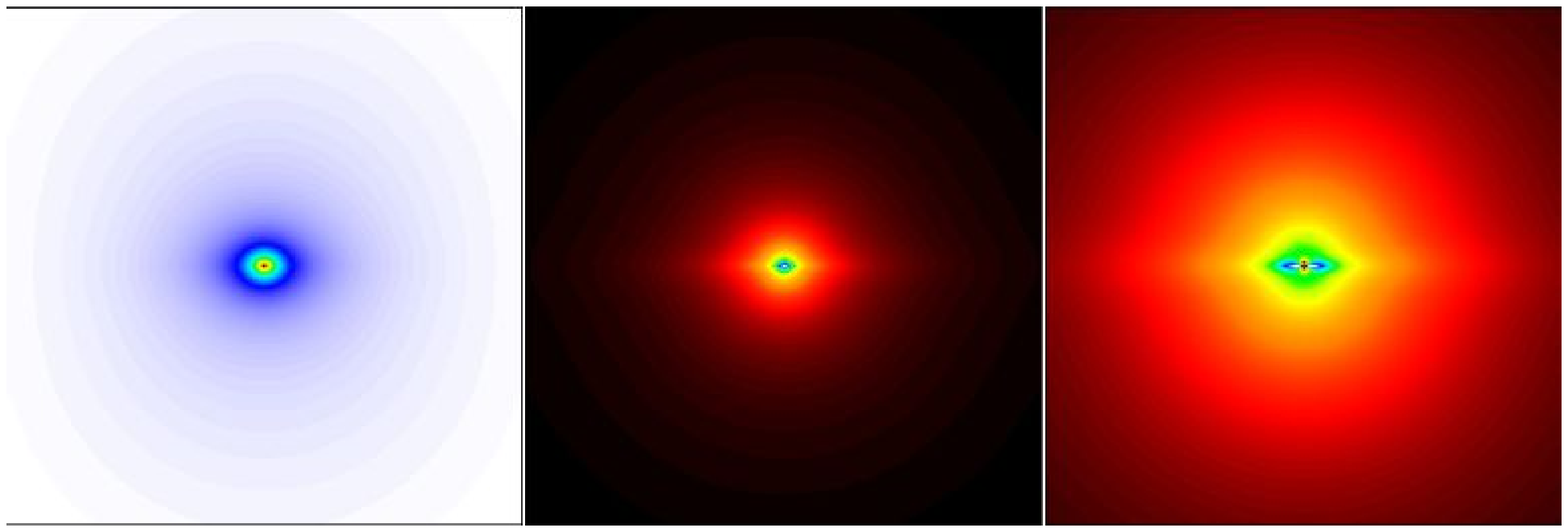}
\caption{A lens model for an almost edge-on spiral galaxy: shown are
  the modulus of the deflection angle (left), the convergence
  (centre), and the absolute value of the shear (right).}
\label{fig:5}
\end{figure*}

As an example, consider a lens model for a spiral galaxy, consisting
of a spherical halo and a flat disk seen almost edge-on (Bartelmann \&
Loeb 1998). The deflection-angle field of such a lens can be given
analytically (cf.~Keeton \& Kochanek 1998). Convergence and total
shear $(\gamma_1+\gamma_2)^{1/2}$ as determined by numerical
differentiation are shown together with the modulus of the deflection
angle in Fig.~\ref{fig:5}.

\begin{figure}[ht]
  \includegraphics[width=0.49\hsize]{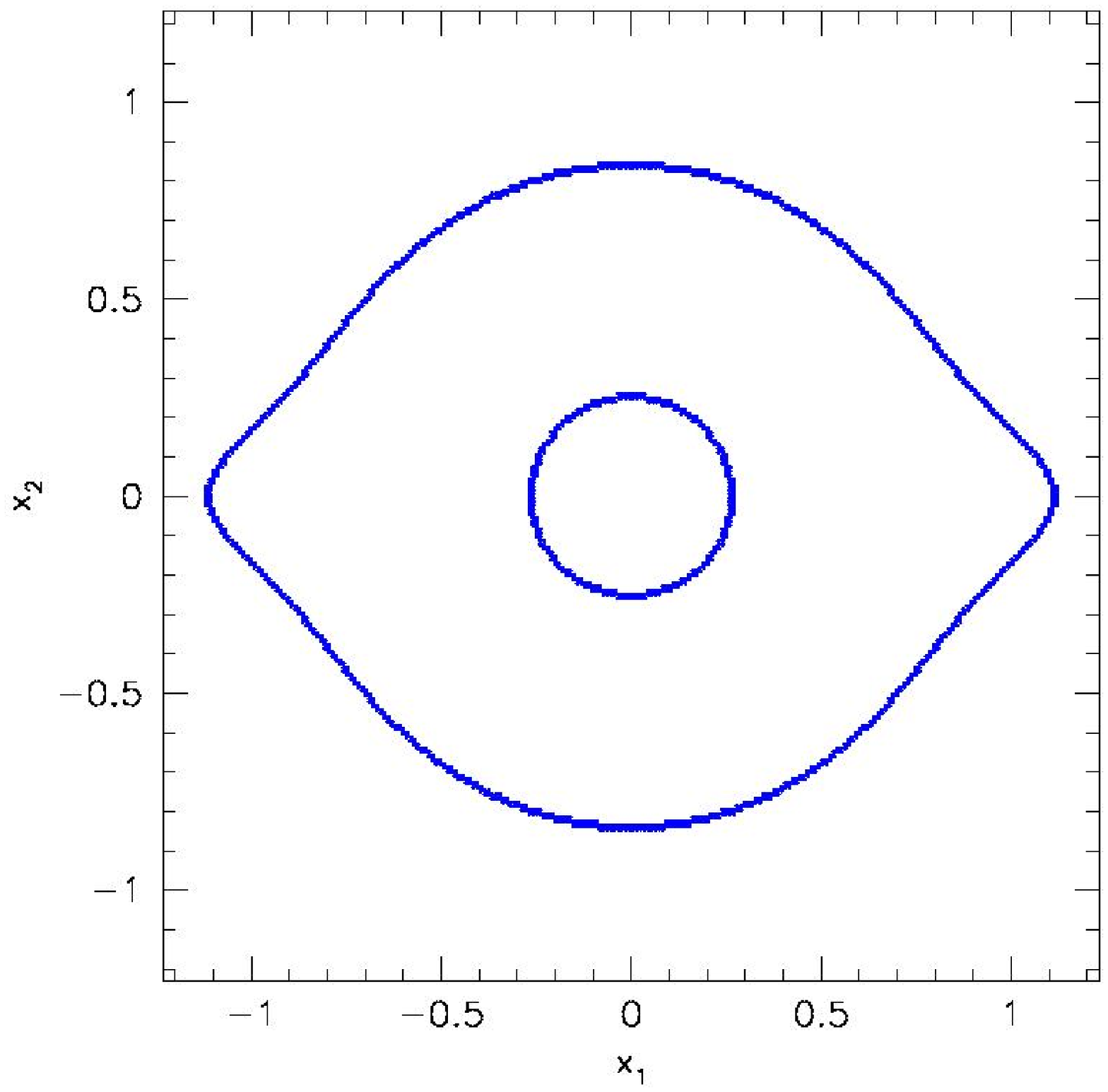}\hfill
  \includegraphics[width=0.49\hsize]{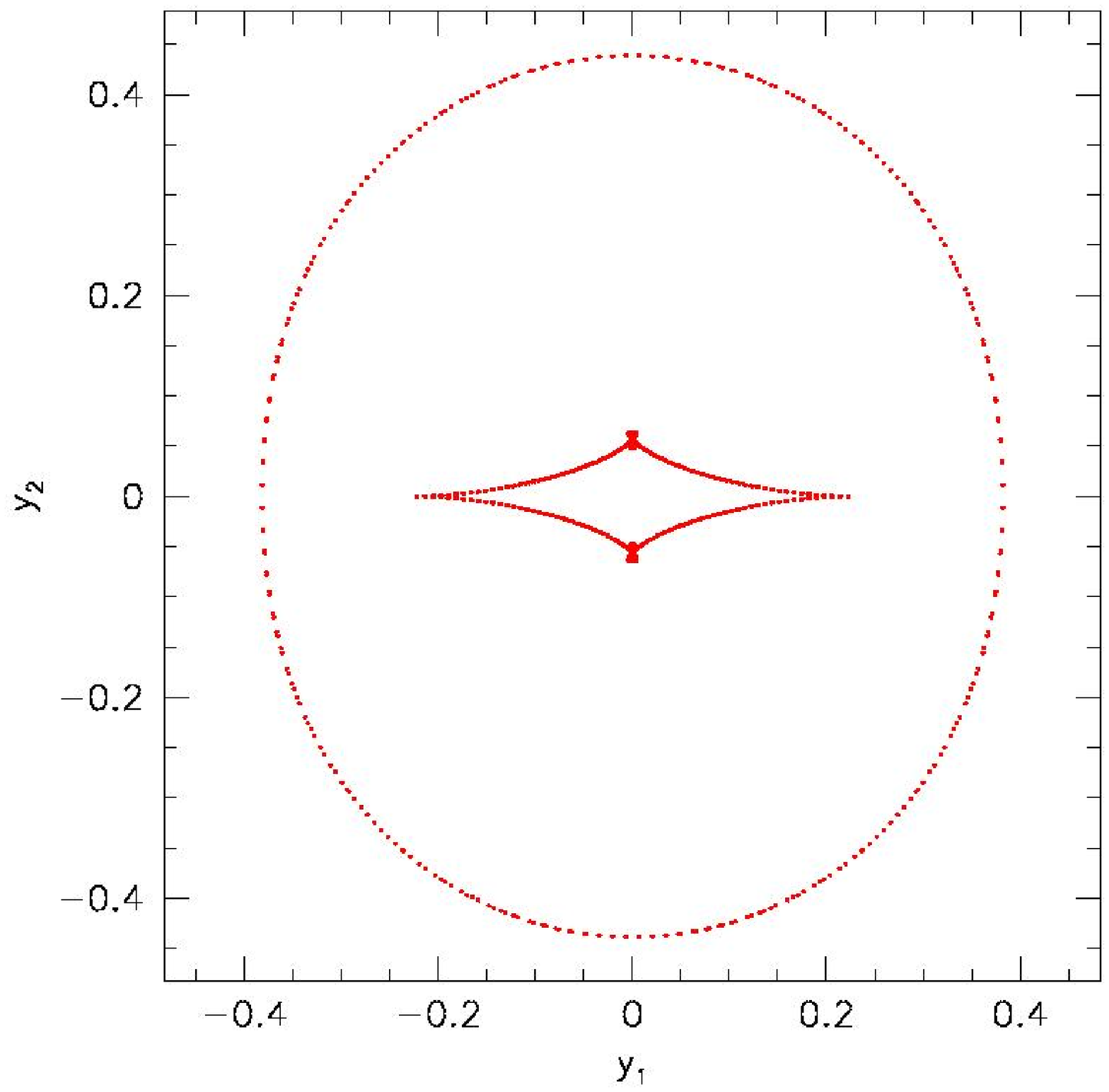}
\caption{Critical curves (left) and caustics (right) of the
  spiral-galaxy lens model illustrated in Fig.~\ref{fig:5}.}
\label{fig:6}
\end{figure}

The critical curves and caustics of that lens model as determined with
the method described above are shown in Fig.~\ref{fig:6}.

\subsection{Adaptive Source Grids}

One of the most prominent goals of gravitational lensing studies with
individual strong lenses is to determine the imaging statistics of a
given lens model, for example the abundance of highly magnified
events, the occurrence of multiple imaging with the images satisfying
certain conditions, and the like. This is done in principle by
distributing many sources across the source plane, imaging them as
described before, and determining the image properties. However, such
events are rare. If one were to cover the entire source plane with a
regular grid of sources, this grid would have to have a very high
resolution for rare events to be reliably found. In turn, most of the
sources probed would produce images failing the criteria imposed, so
by far the largest fraction of the CPU time used would be wasted.

This situation calls for adaptive grids. We know in advance that any
strongly lensed image will occur near a critical curve, or any
strongly lensed source near a caustic. It is those sources that we
need to treat in detail, while those far from caustic curves are
usually only required to normalise the statistics properly.

One approach for defining an adaptive grid, and there may be others
more suitable for a particular lensing situation, proceeds as
follows. Again, we assume that we know the deflection angle of the
lens, either because it was provided numerically or because it is
described by a known analytic formula. Then, we saw in the preceding
subsection how grid points can easily be identified which are close
to a critical curve in the lens plane, or a caustic curve in the
source plane.

In order to save computational time, the source plane is first covered
with a coarse grid. This grid should obviously be fine enough for the
caustics to be properly resolved; for instance, it must not be so
coarse that the typically two types of caustic curve, the radial and
the tangential one, are closer than a few times the grid separation.

Next, those points on that coarse grid are identified and saved which
are next to a caustic curve. This can, for instance, be done by
masking, i.e.~by attaching a logical variable to all grid points and
setting it to either \emph{true} or \emph{false} depending on whether
it is or is not next to a critical curve. One can then cover the
source plane with a grid whose resolution is doubled in both
dimensions, and keep only those points which are identical with, or
surrounded by, points of the coarse grid which were masked in the
preceding step. This procedure can be repeated as often as desired,
i.e.~until the finest grid level reaches the ultimately required
resolution. Note that it is not the grids and their masks which need
to be saved, but only the coordinates of those grid points which are
either part of the coarse initial grid, or whose logical mask values
are \emph{true}. That way, lists of source positions can be
constructed which are to be probed later for the images they give rise
to.

Naturally, this can only be a basic recipe which needs to be adapted
to the situation at hand. For instance, the condition that grid points
need to be next to a caustic can be replaced by the condition that the
absolute value of the Jacobian determinant be less than a given
threshold which can be lowered at each step of grid refinement. Such a
criterion would naturally increase the grid resolution near such
grid positions where sources are certain to be highly magnified.

Of course, if statistics is the ultimate goal, one has to take into
account that sources near caustic curves were positioned such as to
have an unfair advantage over sources far from caustics. Since we have
chosen to double the grid resolution at each refinement step, each
source on a refined grid represents only a quarter of the area on the
source plane represented by a source on the next coarser
grid. Assigning a statistical weight of unity to the sources on the
finest grid, the weight must quadruple for each coarser level. If the
grid was refined $N$ times, the weight of sources on the coarsest grid
is thus $w_i=2^{2N}$. Each source is assigned a statistical weight
$w_i$ in that way, and counts $w_i$ times in the final statistical
evaluation.

The left panel in Fig.~7 shows the source locations chosen for
evaluating image statistics of the spiral lens model illustrated in
Fig.~\ref{fig:6}.

\begin{figure*}[ht]
  \includegraphics[width=0.49\hsize]{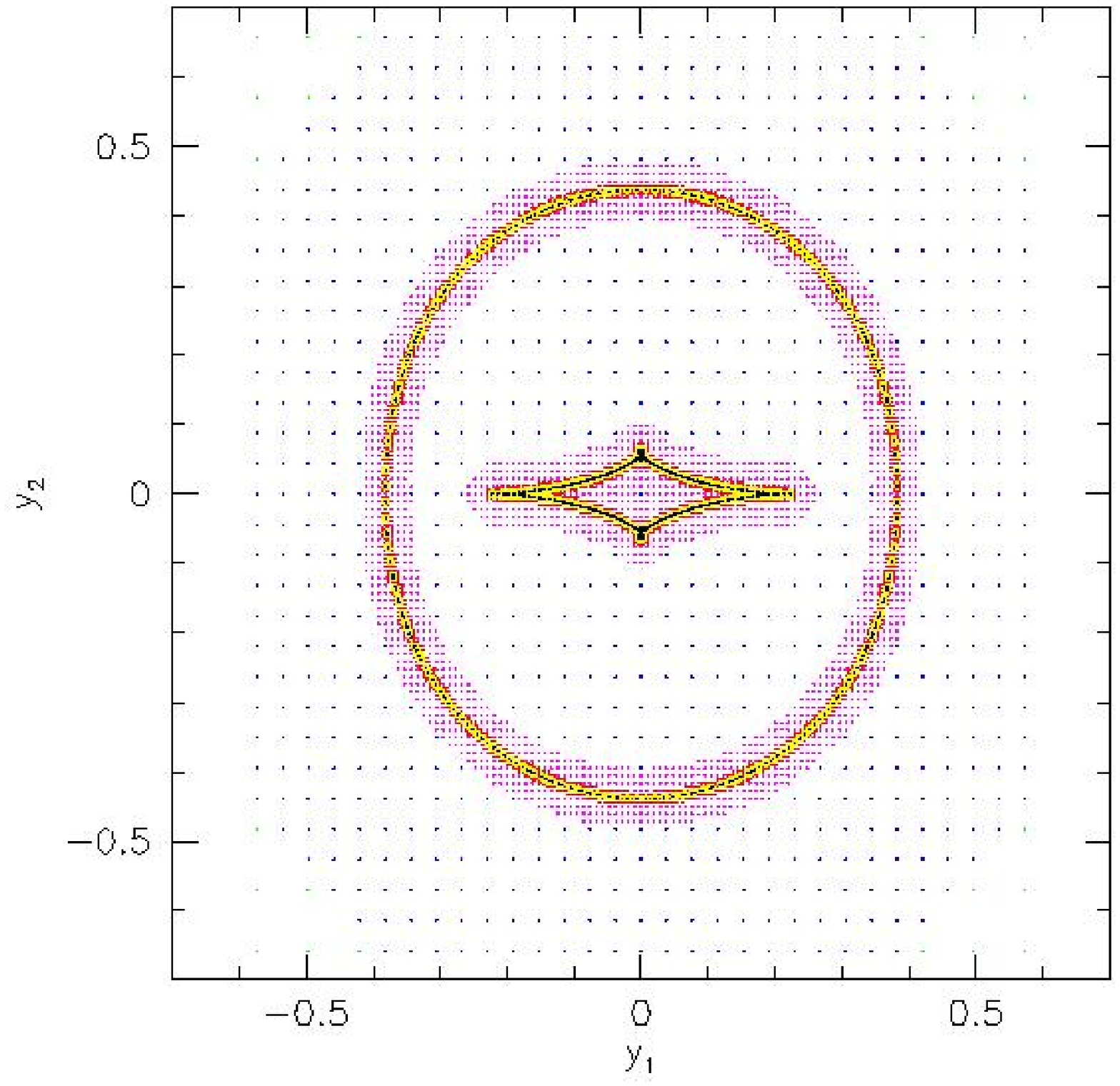}\hfill
  \includegraphics[width=0.49\hsize]{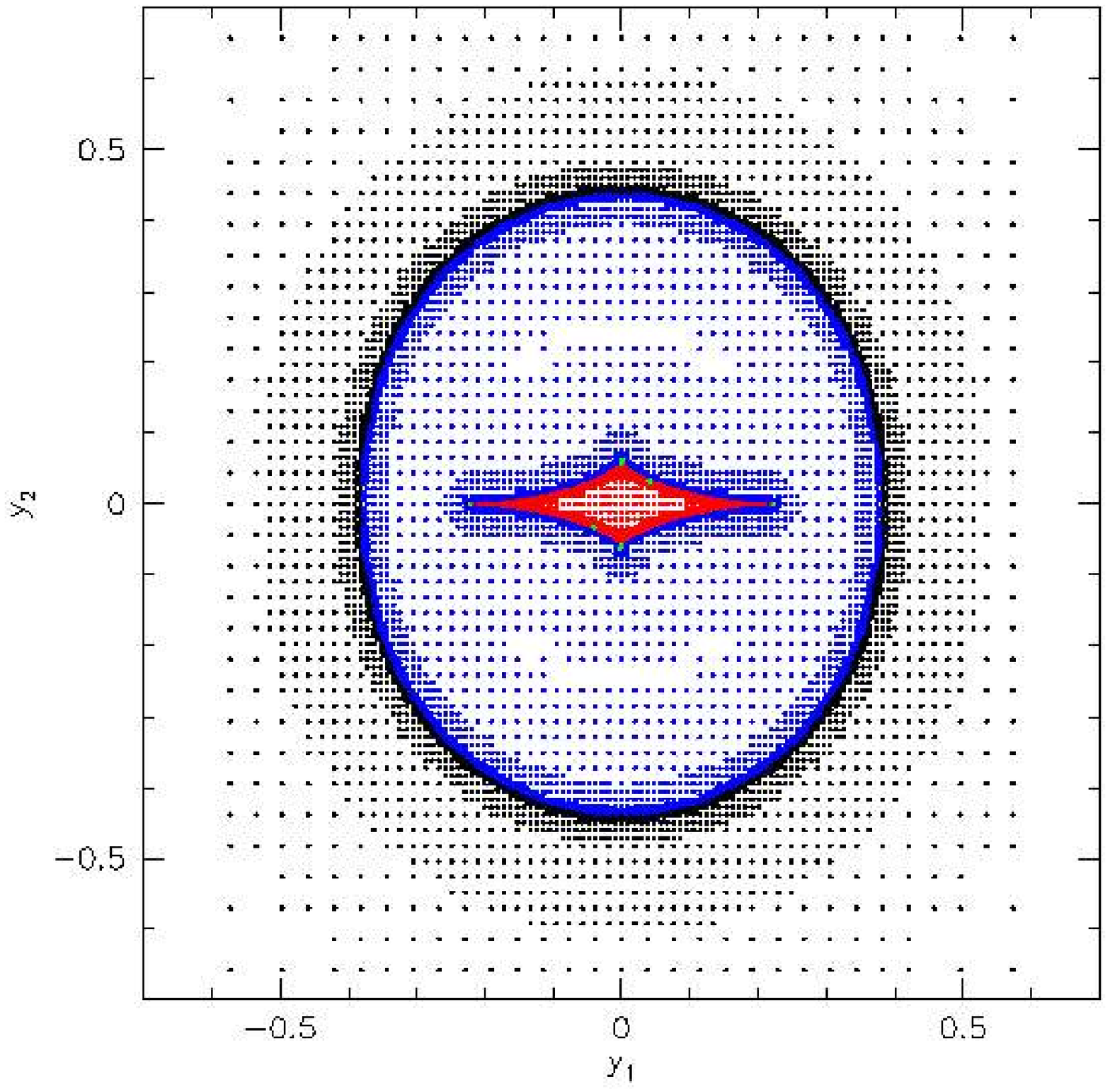}
\caption{\emph{Left panel:} source positions placed on a multiply
  refined grid in the source plane. Caustic points are
  black. Obviously, the source plane is best sampled near
  caustics. \emph{Right panel:} Number of images found for sources
  placed at the positions shown in the left panel.}
\label{fig:7}
\end{figure*}

\subsection{Finding Images}

The principle of finding the images of a given source is simple: Given
a source at position $\vec y_\mathrm{s}$, find those grid points $\vec
x_{ij}$ on the lens plane which are mapped sufficiently close to $\vec
y_\mathrm{s}$, i.e.~whose mapped points $\vec y_{ij}$ are within a
specified distance from $\vec y_\mathrm{s}$.

The problem with this approach is that a square-shaped or rectangular
grid cell from the image plane is mapped onto a distorted figure in
the source plane. In most cases, this figure will be a parallelogram,
but in rare cases, opposing corners of the original rectangle may even
be interchanged on the source plane. How can it then be decided
whether a given point in the source plane is inside or outside the
mapped grid cell, or in other words, whether the image of the given
source falls within that particular grid cell on the lens plane?

The solution is to split each grid cell in the lens plane into two
triangles, because a mapped triangle always remains a triangle, which
always has a well-defined interiour (cf.~Schneider et al.~1992).

\begin{figure}[ht]
  \includegraphics[width=\hsize]{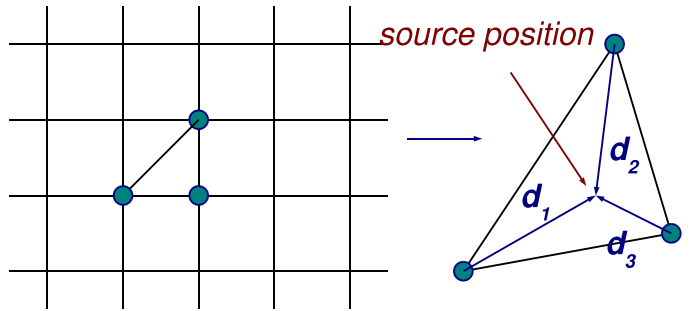}
\caption{Illustration of the technique for finding images described in
  the text. Grid cells in the lens plane are split into triangles
  (left panel), which have a well-defined interior after being mapped
  back onto the source plane (right panel). This would not necessarily
  be the case for rectangular grid cells. A source is contained by a
  triangle if all mixed cross products $\vec d_i\times\vec d_j$ for
  the shown vectors $\vec d_i$ are positive.}
\label{fig:8}
\end{figure}

Consider Fig.~\ref{fig:8}. The three grid points marked on the lens
plane in the left panel of the figure are mapped to the distorted
triangle shown on the right panel, which contains the source
position. Call $\vec d_{1,2,3}$ the three vectors from the mapped
triangle's corners towards the source position. It can be shown that
the source is inside the mapped triangle if the three vector products
\begin{equation}
  \vec d_1\times\vec d_2\;,\;\vec d_1\times\vec d_3\;,\;
  \vec d_2\times\vec d_3
\label{eq:14}
\end{equation}
are all positive, with the vector product in two dimensions being
defined as
\begin{equation}
  \vec a\times\vec b\equiv a_1b_2-a_2b_1\;.
\label{eq:15}
\end{equation}

One straightforward way to verify this condition is to convince one's
self that the source point is inside the triangle if all vectors $\vec
d_i$ point within the angles spanned by the adjacent sides of the
triangle, and that this condition translates to Eq.~(\ref{eq:14})
above.

This algorithm for finding images works well as long as the separation
between images is larger than the size of the grid cells in the lens
plane. Very close images can be contained within the same grid cell,
in which case the algorithm would find only one. Of course, this
potential problem can be remedied by increasing the grid resolution,
but then a very large number of grid cells would have to be checked
in vain for containing an image.

Again, a viable solution uses adaptive grids. One can start with a
coarse grid on the lens plane. Searching for images on that coarse
grid will almost certainly not yield all images of a multiply imaged
source, but those missed will be closer than the grid separation to
those found. Then, those grid cells containing images can individually
be covered with a highly resolved grid, and the image search repeated
on those sub-grids. Hence, the first step represents a coarse scan of
the lens plane for grid cells containing at least one image, and the
second step scans only those regions on the lens plane in detail where
images are sure to be found. If needed, further sub-grids can be
similarly nested.

Of course, even though this procedure is highly adaptive and
efficient, it always has a remaining resolution limit, and images
closer than that will not be resolved. It is then important to adapt
the resolution of the finest sub-grid to the situation at hand, for
instance such that remaining unresolved images would neither be
resolved by observations. The right panel in Fig.~\ref{fig:7}
illustrates the result of an adaptive image search for all sources at
the positions shown in the figure's left panel. Colours denote image
numbers: Black means one image, blue three, and red five, while green
shows source positions for which an even number of images has been
found, in contradiction to the necessarily odd image number produced
by non-singular lenses. Such events are rare, but they do occur
because of the finite resolution limit of the algorithm applied.

\begin{figure}[ht]
  \includegraphics[width=\hsize]{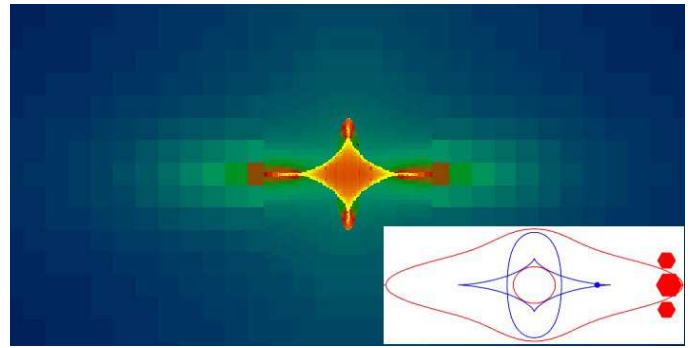}
\caption{The colour encodes the total magnification of a point source
  lensed by an almost edge-on spiral galaxy; blue means a
  magnification near unity, yellow means very high magnification. The
  adaptive resolution of grid cells on the source plane is clearly
  visible. The size of the grid cells decreases substantially towards
  regions of high magnification. The inserted panel shows caustics and
  critical curves of the same lens (blue and red, respectively), a
  source position close to the right-hand ``naked'' cusp, and the
  three images as red hexagons, whose size logarithmically encodes
  their magnification.}
\label{fig:9}
\end{figure}

Figure~\ref{fig:9} gives an example for possible results of that
adaptive technique for finding images. Colour-coded is the total
magnification of point sources in the source plane behind the almost
edge-on spiral lensing galaxy introduced above. The increasing spatial
resolution towards the caustic curves is evident. The panel inserted
into the figure shows caustics (blue) and critical curves (red) of the
lens, the source position as a blue dot just inside the right-hand
``naked'' cusp, and the three images as red hexagons whose size
logarithmically encodes the image magnification.

\subsection{Asymmetric Lenses}

So far, we have used a model for a spiral galaxy as an example for a
complex lens whose properties need to be determined
numerically. Despite its complexity, the model is still highly
symmetric; and what is more, its deflection angle is given as an
analytic formula. Sources were so far assumed to be point-like.

Let us now increase the level of complexity and use a numerically
simulated galaxy cluster to gravitationally lens extended
sources. Again, we assume the deflection angle to be given and
postpone the question as to how it can be determined from an $N$-body
simulation.

\begin{figure*}[ht]
  \includegraphics[width=0.49\hsize]{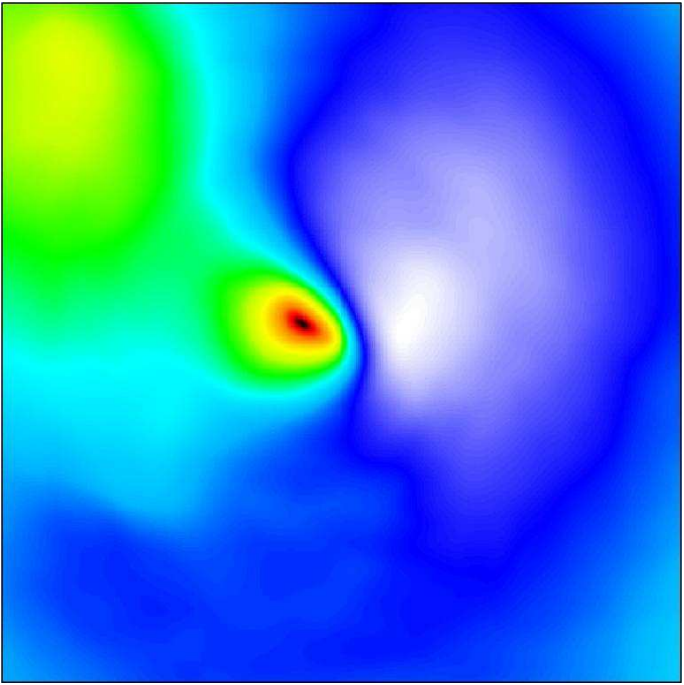}\hfill
  \includegraphics[width=0.49\hsize]{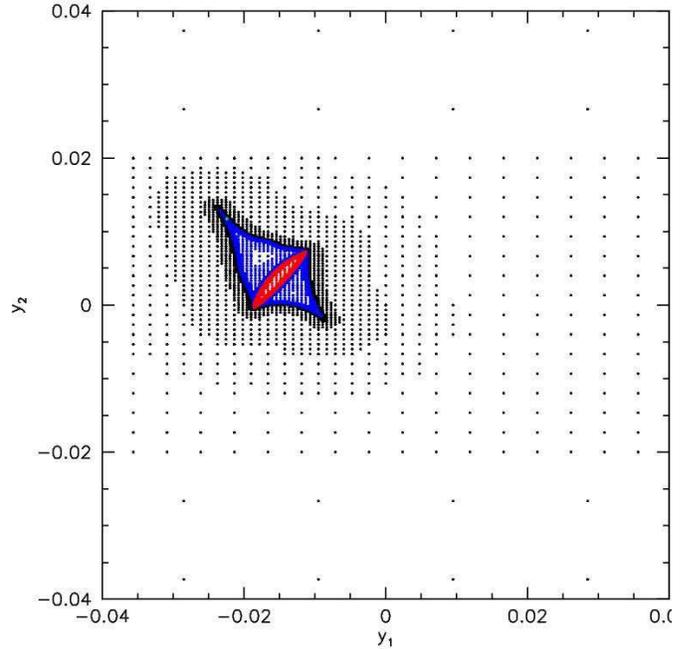}
\caption{The colour plot in the left panel shows the modulus of the
  deflection-angle field in the lens plane of a numerically simulated
  galaxy cluster. The right panel shows how sources are adaptively
  placed on the source plane (dots), and how many images these sources
  have; one (black), three (blue) or five (red). The boundaries
  between the colours mark the caustic structure.}
\label{fig:10}
\end{figure*}

All techniques described above for computing convergence and shear
from the deflection angle, for finding critical curves and caustics,
for placing sources on an adaptive grid, and for finding images within
grid cells split into triangles remain valid
unchanged. Figure~\ref{fig:10} shows an example.

The modulus of the cluster's deflection angle is shown as the colour
plot in the left panel. The right panel shows a section of the source
plane with the dots marking source positions, and their colour
illustrating the image number. Black, blue and red means one, three,
or five images, respectively. The caustic structures can clearly be
identified as the boundaries between black and blue and between blue
and red, respectively.

\subsection{Imaging Extended Sources}

Extended sources can be described in a variety of ways. What follows
is a simple description for elliptical sources, but alternative source
models can easily be constructed along similar lines. We assume that
source positions $\vec y_\mathrm{s}$ have already been found,
preferentially on an adaptive grid as described before. Also, we need
to be sure that the grid resolution in the source plane is
sufficiently high as to resolve the smallest sources to be considered.

Elliptical sources are described by three more parameters, viz.~their
size, their ellipticity, and their position angle $\phi$. Let us
describe the ellipticity by $e=b/a$, with $a$ and $b$ being the
semi-major and semi-minor axes of the ellipse, respectively. Finally,
we introduce an effective radius $r$ by demanding that a circle of
radius $r$ have the same area as the ellipse, hence $r=\sqrt{ab}$. By
rotating by an angle $\phi$ an ellipse centred on the coordinate
origin whose axes are aligned with the coordinate axes, it can
straightforwardly be shown that a grid point $\vec y_{ij}$ is enclosed
by the ellipse if the condition
\begin{eqnarray}
  \cos^2\phi\left(\frac{\delta y_1^2}{e}+e\delta y_2^2\right)&+&
  \sin^2\phi\left(\frac{\delta y_2^2}{e}+e\delta y_1^2\right)
  \nonumber\\ &+&
  2\delta y_1\delta y_2\sin\phi\cos\phi\left(\frac{1}{e}-e\right)
  \le r^2
\label{eq:16}
\end{eqnarray}
is satisfied, where $\delta\vec y\equiv\vec y_{ij}-\vec
y_\mathrm{s}$. If the grid point $\vec x_{ij}$, whose image in the
source plane is $\vec y_{ij}$, satisfies Eq.~(\ref{eq:16}), the image
point $\vec x_{ij}$ is part of the source, and the image can be
constructed by assigning the source's surface brightness at $\vec
y_{ij}$ to the image point $\vec x_{ij}$. By mapping the entire lens
plane onto the source plane and checking Eq.~(\ref{eq:16}) for each
individual imaged grid point $\vec y_{ij}$, all image points belonging
to the given source can be identified.

It is often desired for statistical purposes to automatically
characterise a large number of images. An example is the determination
of cross sections for the formation of large gravitational arcs by a
numerically simulated galaxy cluster, for which a large number of
sources need to be imaged and the image properties automatically
quantified to search for the rare ``giant'' arcs. Most of the methods
described here have been introduced and used extensively e.g.~by
Bartelmann \& Weiss (1994), Bartelmann et al.~(1995, 1998), Meneghetti
et al.~(2000, 2001); see also the contribution by Massimo Meneghetti
to this volume.

A source may have multiple images, thus the point sets in the lens
plane found by imaging extended sources need not be connected. The
first step is therefore to group the image points into images. This
can be done with a variant of the classical friends-of-friends
algorithm: Pick one arbitrary point out of any given set of image
points and search for another image point which is at most $\sqrt{2}h$
grid units away from the first point; $h$ is the grid size in the lens
plane. If there is such a point, it is called a ``friend'' and grouped
into the same image as the first point. Now take the ``friend'' and
repeat until no further ``friends'' can be found and the image is
complete. If more image points are left on the lens plane, pick one of
those and repeat the process until all image points have been
grouped. If the image is large enough, and the grid resolution on the
lens plane is high enough for the image to consist of many points, the
image magnification is simply the ratio between the numbers of pixels
covered by an image and the number of pixels covered by the source.

Once all image points belonging to a single image have been
identified, it is often useful to determine the boundary points of
that image, e.g.~by identifying those points inside an image which
have a neighbour outside the image. By suitably ordering the boundary
points, a boundary line can be found whose length can be measured and
used in further steps of the automatic image classification. Next, the
curvature of the image can be found by first identifying the image
point which of the source centre, then search for the boundary point
most distant from the so-defined image centre, and finally searching
for the boundary point most distant from the first boundary
point. These three points uniquely define a circle whose radius can be
used as an approximation for the arc radius. And so on, you get the
drift: Once image points are grouped into individual images and
boundary curves have been determined, images can be classified by
adapting elementary geometrical figures to them.

\subsection{Deflection Angles of Asymmetric Lenses}

So far we have assumed to be given the deflection angle either as an
analytic expression or as two two-dimensional arrays of numbers giving
its two components as a function of position in the lens plane. We now
need to describe methods for obtaining the deflection angle of a
numerically simulated lens.

The first issue to be discussed is the spatial resolution. Since the
simulated lens is composed of discrete particles which represent a
smooth mass distribution in reality, the deflection angle must not be
computed by simply summing up the deflection angles of the individual
particles: The result would be a collection of microlenses rather than
a single macrolens, having many spurious and undesired imaging
features.

Rather, the collection of particles has to be projected onto a lens
plane, on which it needs to be smoothed in some way. We will return
later to the issue of how particles should be sorted into grid
cells. An important point to be addressed before is how large the grid
cells should be chosen. They should be small enough for important
features of the lens to remain identifiable; they should be large
enough for the surface density to lose the ``graininess'' due to its
being composed of individual particles, and they should be large
enough so that Poisson errors are smaller than a certain threshold. If
the number of particles per grid cell is $nh^2$, its Poisson
fluctuation is $\sqrt{nh^2}$, thus the discreteness of the particles
gives rise to fluctuations in the surface-mass density. Demanding that
the relative fluctuations of the density should be smaller than
$\epsilon\ll1$, the cell size $h$ has to be chosen such as to satisfy
$(nh^2)^{-1/2}\le\epsilon$. It is impossible to give a general rule
applicable to the majority of lensing situations, but it is clear that
resolution, smoothing and particle noise have to be carefully balanced
by choosing the grid cell size appropriately.

Assigning particle masses to grid points in order to obtain a smooth
density distribution is an art of its own (cf.~Hockney \& Eastwood
1988). In principle, the particle mass could simply be attributed to
the single grid point next to its position. This ``nearest grid
point'' (NGP) method is appropriate for particles near the centre of a
cell, but particles near cell boundaries should be attributed to the
cell and its neighbour(s) in order to avoid boundary effects like
density discontinuities. Numerous schemes for interpolating particles
across cells have been proposed. They are generally of the form
\begin{equation}
  Q(\vec x)=\sum_i\,W\left(\vec x-\vec x_i\right)\,
  Q(\vec x_i)\;,
\label{eq:17}
\end{equation}
where $Q$ is the quantity to be interpolated onto a point $\vec x$,
e.g.~the particle mass, the sum extends over all particles
sufficiently close to the point of interest $\vec x$, and $W(\vec
x-\vec x_i)$ is a smoothing or interpolation kernel depending on the
separation vector between the particle position $\vec x_i$ and $\vec
x$. The kernel is decomposed into three factors
directions,
\begin{equation}
  W(\delta\vec x)=w(\delta x_1)w(\delta x_2)w(\delta x_3)\;,
\label{eq:17a}
\end{equation}
one for each dimension, the $i$-th of which depends only on the
$i$-component of the separation vector. Interpolation methods can now
be classified according to the kernel factors $w(\delta x)$ and their
width.

\begin{figure}[ht]
  \includegraphics[width=0.49\hsize]{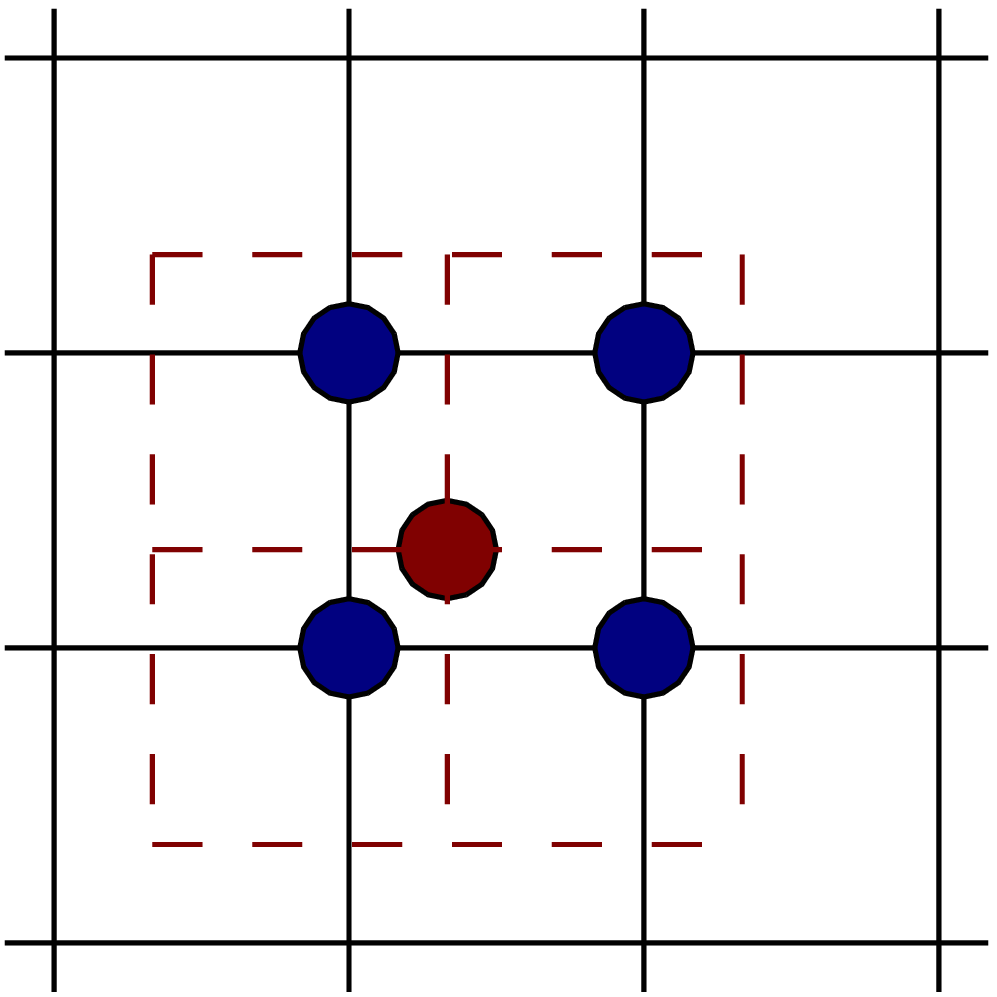}\hfill
  \includegraphics[width=0.49\hsize]{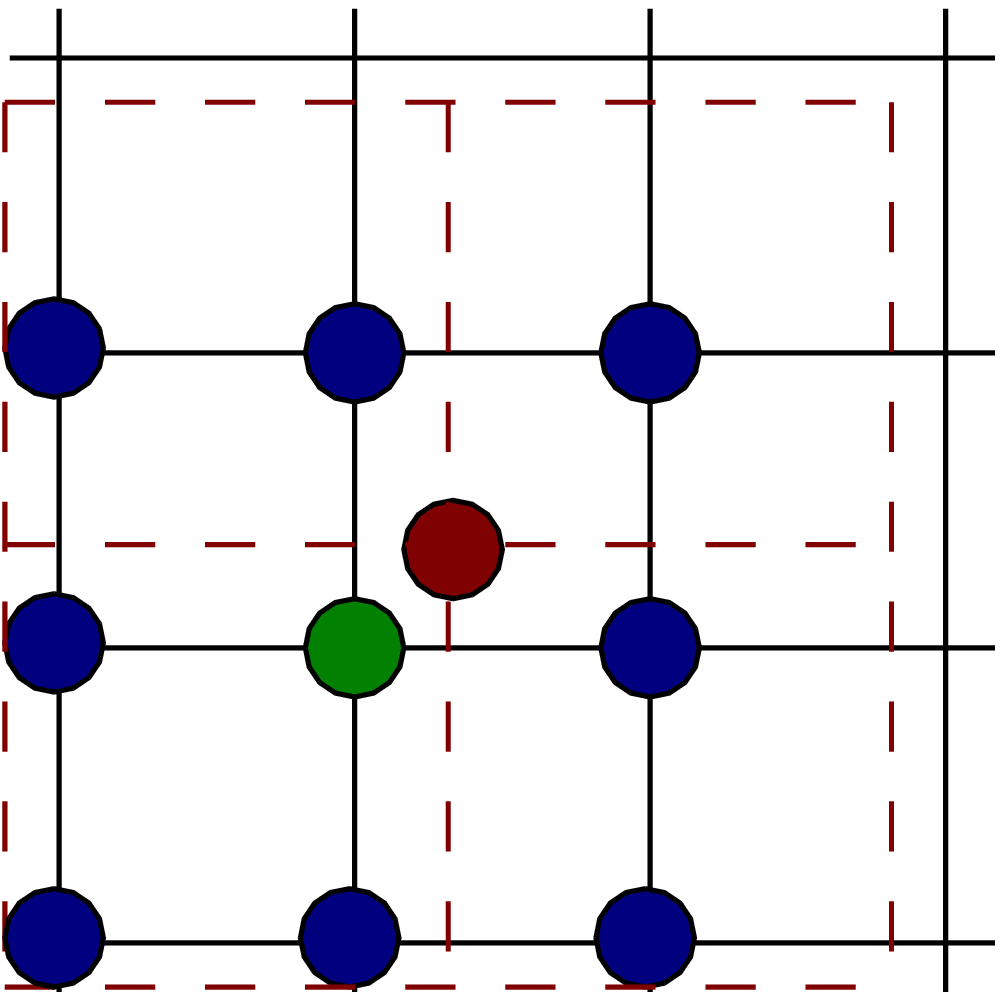}
\caption{The ``cloud-in-cell'' (left panel) and ``triangular shaped
  cloud'' (right panel) interpolation schemes are illustrated
  here. The (projected) particle position is marked red, the grid
  points to which the particle mass is assigned are marked blue and
  green. The CIC and TSC schemes assign the particle mass to the eight
  and 27 nearest neighbours, respectively (in three dimensions).}
\label{fig:11}
\end{figure}

The ``cloud-in-cell'' (CIC) scheme uses the kernel factors
\begin{equation}
  w_\mathrm{CIC}(\delta x)=\left\{\begin{array}{ll}
    1-|\delta x|/h & \mbox{for}\quad |\delta x|<h \\
    0              & \mbox{otherwise}
			   \end{array}\right.\;,
\label{eq:18}
\end{equation}
which implies that the particle is distributed over the four nearest
grid points. A more elaborate scheme is the ``triangular shaped
cloud'' (TSC) method, which uses the kernel factors
\begin{equation}
  w_\mathrm{TSC}(x)=\left\{\begin{array}{ll}
    3/4-\delta x^2/h^2     & \mbox{for}\quad |\delta x|\le h/2 \\
    (3/2-|\delta x|/h)^2/2 & \mbox{for}\quad h/2\le|\delta x|<3h/2 \\
    0                     & \mbox{otherwise}
			   \end{array}\right.\;.
\label{eq:19}
\end{equation}
The CIC and TSC interpolation schemes are illustrated for two
dimensions in Fig.~\ref{fig:11}. For all schemes, the kernel has to be
normalised such that all particle mass fractions add up to unity.

Suppose now we have obtained the surface mass density on a grid
$\kappa_{ij}=\kappa(\vec x_{ij})$, then the deflection angle can most
straightforwardly be determined by direct summation as
\begin{equation}
  \vec\alpha_{ij}=\frac{1}{\pi}\,\sum_{kl}\,\kappa_{kl}\,
  \frac{\vec x_{ij}-\vec x_{kl}}
       {\left|\vec x_{ij}-\vec x_{kl}\right|^2}\;.
\label{eq:20}
\end{equation}
Depending on the number of grid cells, the direct summation can be
prohibitively slow. In many circumstances of astrophysical interest,
fast-Fourier techniques can then be applied. In order to see how this
works, note that the deflection angle can be written as a convolution
of the convergence $\kappa(\vec x)$ with a kernel
\begin{equation}
  \vec K(\vec x)=\frac{1}{\pi}\,\frac{\vec x}{\left|\vec x\right|^2}\;.
\label{eq:21}
\end{equation}
This allows the Fourier convolution theorem to be applied, which holds
that the Fourier transform of a convolution is the product of the
Fourier transforms of the functions to be convolved, hence
\begin{equation}
  \hat{\vec\alpha}(\vec k)=\hat\kappa(\vec k)\hat{\vec K}(\vec k)\;.
\label{eq:22}
\end{equation}
The Fourier transform of the kernel $\vec K$ can be determined and
tabulated once. Using fast-Fourier techniques to determine the Fourier
transform of the convergence $\hat\kappa(\vec k)$ requires the
convergence to be periodic on the lens plane. In many cases, this can
be safely assumed or arranged. Often, lens planes are constructed from
large-scale $N$-body simulations which have periodic boundary
conditions by design, or the lens is an isolated object like a galaxy
cluster, which can be surrounded by a sufficiently large field for the
convergence to drop near zero everywhere around the edges of the
field. Fast-Fourier methods speed up the computation of the deflection
angle considerably.

If necessary, derivatives of the deflection angle field can also be
determined in Fourier space. Once the convergence has been Fourier
transformed, one can employ the two-dimensional Poisson equation to
compute the Fourier transform of the lensing potential,
\begin{equation}
  \hat\psi=-\frac{2}{k^2}\,\hat\kappa\;,
\label{eq:23}
\end{equation}
from which the Fourier transforms of the deflection angle and the
shear components can easily be determined,
\begin{equation}
  \hat{\vec\alpha}=-\mathrm{i}\,\vec k\,\hat\psi\;,\quad
  \hat\gamma_1=-\frac{1}{2}\left(k_1^2-k_2^2\right)\,\hat\psi\;,\quad
  \hat\gamma_2=-k_1k_2\,\hat\psi\;.
\label{eq:24}
\end{equation}
Relations like those and the exploitation of fast-Fourier methods are
particularly relevant for simulating gravitational lensing by
large-scale structures.

\section{Lensing by Large-Scale Structures}

\subsection{Resolution Issues}

Obviously, the thin-lens approximation that we have been using so far
breaks down if one wishes to study gravitational lensing by
large-scale structures. The solution then is to cover the complete
cosmic volume whose lensing effects one wants to simulate with
simulation boxes stacked along the line-of-sight, to project suitable
slices on individual lens planes, and to use multiple lens-plane
theory for describing light propagation.

\begin{figure*}[ht]
  \includegraphics[width=0.49\hsize]{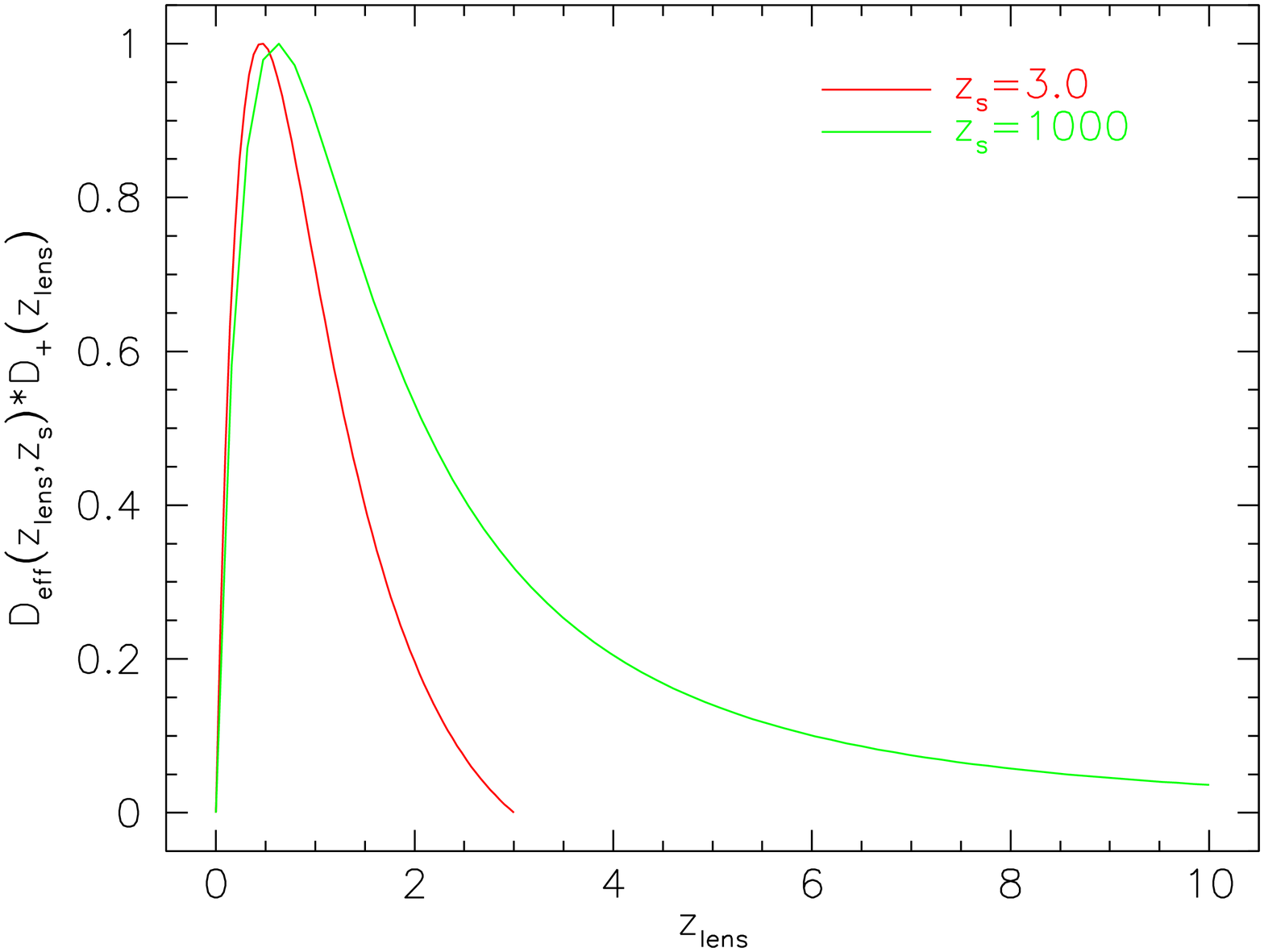}\hfill
  \includegraphics[width=0.49\hsize]{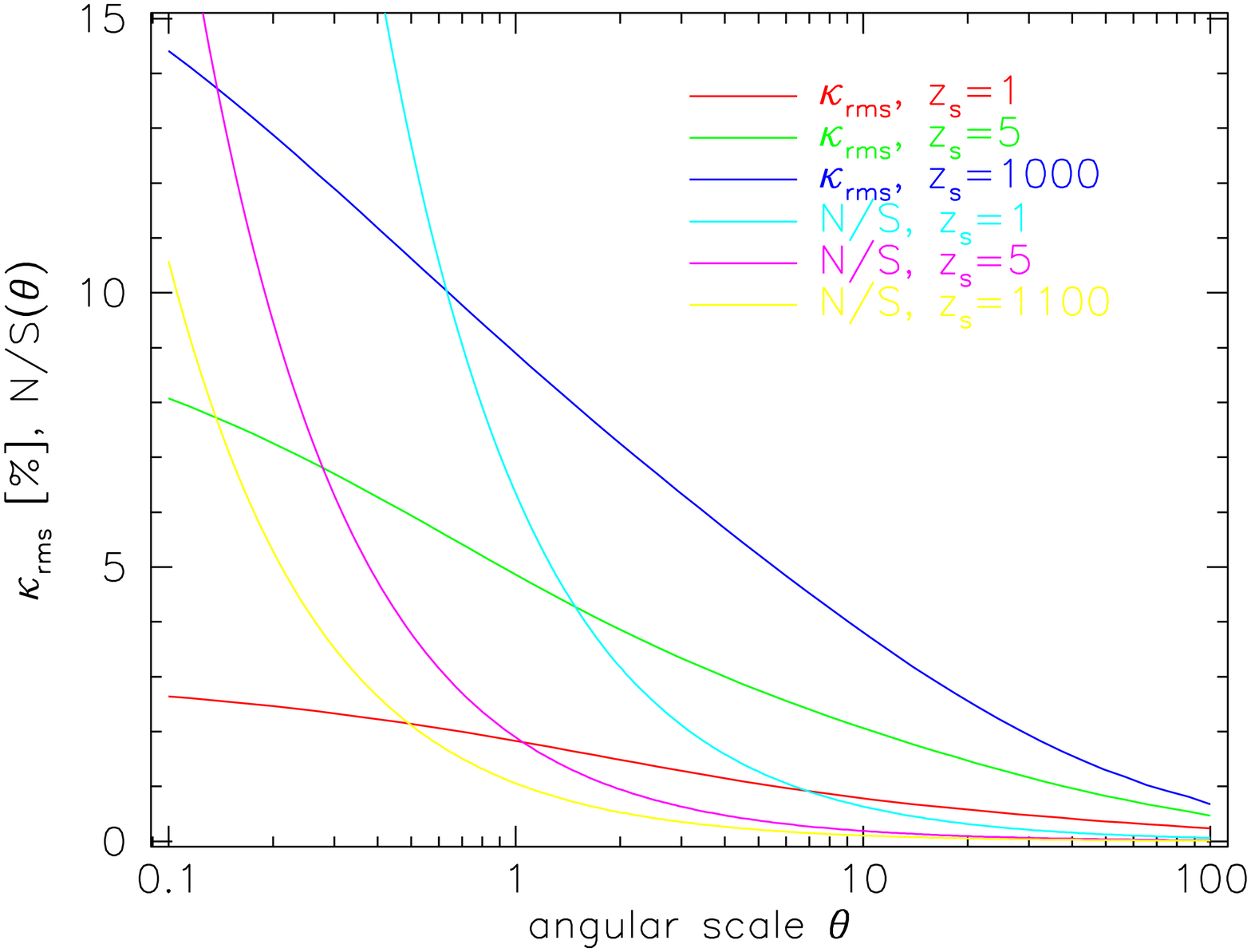}
\caption{\emph{Left panel}: The product of lensing efficiency function
  times the linear growth factor for density perturbations is shown
  for two different source redshifts, $z_\mathrm{s}=3$ and
  $z_\mathrm{s}=1000$, respectively, the latter being relevant for
  gravitational lensing of the cosmic microwave background. The growth
  factor skews the geometrical lensing efficiency towards lower
  redshifts. At redshift 5, the combined efficiency function drops to
  10\% of its peak value for $z_\mathrm{s}=1000$, implying that by far
  not the complete redshift range up to $z_\mathrm{s}$ needs to be
  covered with lens planes. \emph{Right panel}: The solid curves show
  the \emph{rms} cosmic convergence for sources at three different
  redshifts in per cent, the dashed curves the noise-to-signal ratio
  obtained in an $N$-body experiment with particle mass
  $m_\mathrm{p}=6.8\times10^{10}\,h^{-1}\,M_\odot$. Both types of
  curve are plotted as functions of angular scale.}
\label{fig:12}
\end{figure*}

The multiplicity of lens planes, and the general weakness of lensing
by large-scale structures, make questions of angular and mass
resolution particularly relevant for cosmic lensing. For instance,
lens planes close to the observer are typically poorly resolved
because even small grid cells span a large solid angle near the
observer, and making grid cells smaller is not generally an acceptable
solution because then the number of particles per grid cell becomes
small, and the shot noise possibly unacceptably large. However, lens
planes near the observer are less efficient than lens planes
approximately half-way to the source because the lensing efficiency
function is zero at the observer and source redshifts and peaks in
between. Yet, structures grow over time, thus the lensing efficiency
function is skewed towards lower redshifts because structures are
geometrically less efficient lenses, but their density contrast keeps
growing. By a related argument, sources at very high redshifts do not
require the entire cosmological volume between them and the observer
to be filled with lens planes because lens planes at very high
redshift are geometrically inefficient and have a low density
contrast. The left panel of Fig.~\ref{fig:12} shows two examples for
the lensing efficiency function times the linear growth factor, which
is the relevant quantity combining structure growth with geometrical
efficiency.

Similarly, the effective angular resolution of the simulation is
dominated by the angular resolution of those lens planes near the peak
in the combined efficiency function, i.e.~the product of geometrical
efficiency and linear growth factor.

The shot noise caused by the discretisation of mass into particles is
particularly important for studies of weak lensing by large-scale
structures. Even in absence of density inhomogeneities, shot noise
leads to density fluctuations. They need to be sufficiently smaller
than the signal, i.e.~the convergence fluctuations which cause weak
lensing.

In essence, this requirement also imposes a resolution limit. Suppose
we wish to quantify the weak-lensing signal within a solid angle
$\delta\Omega$. The volume spanned by $\delta\Omega$ within redshifts
$z$ and $z+\d z$ is
\begin{equation}
  \d V(z)=\delta\Omega\,D^2(z)\,
  \left|\frac{\d D_\mathrm{prop}}{\d z}\right|\,\d z\;,
\label{eq:25}
\end{equation}
where $D(z)$ and $D_\mathrm{prop}(z)$ are the angular diameter and
proper distances to redshift $z$. In absence of density
inhomogeneities, this volume element contains $\d N(z)$ particles,
with
\begin{equation}
  \d N(z)=\d V(z)\,\frac{\bar\rho(z)}{m_\mathrm{p}}\;,
\label{eq:26}
\end{equation}
where $\bar\rho(z)$ is the mean matter density at redshift $z$, and
$m_\mathrm{p}$ is the mass of an $N$-body particle in the
simulation. The contribution to the lensing convergence by these
particles has to be weighted by the effective lensing distance,
$D_\mathrm{eff}(z,z_\mathrm{s})$, and by numerical factors. Poisson
fluctuations in the particle number thus cause convergence
fluctuations whose variance is
\begin{equation}
  \delta^2\kappa\propto\int_0^{z_\mathrm{s}}
  \d z\,D_\mathrm{eff}^2(z,z_\mathrm{s})\,\d N(z)\;.
\label{eq:27}
\end{equation}
These fluctuations need to be compared with, and smaller than, the
convergence fluctuations due to large-scale structure, which are
typically of order $\langle\kappa^2\rangle^{1/2}\approx5\%$ for
sources near redshift unity and angular scales of order
$1'$. According to Eqs.~(\ref{eq:25}) through (\ref{eq:27}), the
\emph{rms} shot noise $\langle\delta^2\kappa\rangle^{1/2}$ scales like
$\delta\Omega^{1/2}$, thus the requirement that the signal-to-noise
ratio
\begin{equation}
  \frac{\mathrm{S}}{\mathrm{N}}=\left(
    \frac{\langle\kappa^2\rangle}{\langle\delta^2\kappa\rangle}
  \right)^{1/2}
\label{eq:28}
\end{equation}
exceed a specified threshold translates into a lower limit to the
solid angle $\delta\Omega$ which can reasonably be resolved by the
simulation. The smallness of the \emph{rms} cosmic convergence
$\kappa_\mathrm{rms}=\langle\kappa^2\rangle^{1/2}$ implies that many
particles need to be enclosed by the ``cone'' spanned by
$\delta\Omega$ for the simulation to be reliable. The right panel of
Fig.~\ref{fig:12} shows an example. The \emph{rms} cosmic convergence
in per cent and the \emph{noise-to-signal} ratio are plotted as
functions of angular scale. The noise level was adapted to an $N$-body
simulation with particle mass
$m_\mathrm{p}=6.8\times10^{10}\,h^{-1}\,M_\odot$. The curves show that
the noise-to-signal ratio drops below unity for sources at redshift
$z_\mathrm{s}=1$ only if the angular resolution is lowered to
$\gtrsim5'$, while an angular resolution of $\gtrsim0.8'$ can be
achieved for $z_\mathrm{s}=1000$ (i.e.~for weak lensing of the CMB;
Pfrommer 2002).

\subsection{Multiple Lens-Plane Theory}

Weak lensing by large-scale structures requires the cosmic volume to
be split into multiple lens planes rather than a single one (for
general reference on multiple lens-plane theory, see Schneider et
al.~1992). The lens plane closest to the observer is the image plane
which represents the observer's sky. A light ray piercing the image
plane at a physical coordinate $\vec\xi_1$ is mutiply deflected on $N$
lens planes and finally reaches the source plane at the physical
coordinate
\begin{equation}
  \vec\eta(\vec\xi_1)=\frac{D_\mathrm{s}}{D_1}\vec\xi_1+
  \sum_{i=1}^N\,D_{i\mathrm{s}}\,\vec{\hat\alpha}(\vec\xi_i)\;,
\label{eq:29}
\end{equation}
where the $D_i$ and $D_{i\mathrm{s}}$ are the angular diameter
distances from the observer to the $i$-the lens plane, and from the
$i$-th lens plane to the source, respectively. The light ray passes
the $i$-th plane at $\vec\xi_i$, where it is deflected by
$\vec{\hat\alpha}(\vec\xi_i)$. Similarly, the $\vec\xi_i$ are
determined by
\begin{equation}
  \vec\xi_j(\vec\xi_1)=\frac{D_j}{D_1}\vec\xi_1+
  \sum_{i=1}^{j-1}\,D_{ij}\,\vec{\hat\alpha}(\vec\xi_i)\;,
\label{eq:30}
\end{equation}
where $D_{ij}$ is the angular diameter distance from the $i$-th to the
$j$-th lens plane.

Introducing angular coordinates $\vec\theta_i=\vec\xi_i/D_i$ yields
\begin{equation}
  \vec\theta_j(\vec\theta_1)=\vec\theta_1+
  \sum_{i=1}^{j-1}\,\frac{D_{ij}D_\mathrm{s}}{D_jD_\mathrm{is}}\,
  \vec\alpha(\vec\theta_i)\;,
\label{eq:31}
\end{equation}
where we have introduced the \emph{reduced} deflection angle
$\vec\alpha=(D_{i\mathrm{s}}/D_\mathrm{s})\hat{\vec\alpha}$. We now
define the matrices
\begin{equation}
  \mathcal{A}_i=\frac{\partial\vec\theta_i}{\partial\vec\theta_1}
  \;,\quad
  \mathcal{U}_i=\frac{\partial\vec\alpha_i}{\partial\vec\theta_i}\;.
\label{eq:32}
\end{equation}
Clearly, $\mathcal{A}_i$ is the Jacobian matrix of the lens mapping
between the $i$-th lens plane and the image plane, thus
$\mathcal{A}_N$ is the Jacobian matrix of the mapping between the
source and image planes. The goal is thus to determine $\mathcal{A}_N$
in order to obtain convergence, shear, and magnification for a light
ray starting out into direction $\vec\theta_1$. The ray-tracing
equation (\ref{eq:31}) implies the recursion relation
\begin{equation}
  \mathcal{A}_j=\mathcal{I}-
  \sum_{i=1}^{j-1}\,\frac{D_{ij}D_\mathrm{s}}{D_jD_\mathrm{is}}\,
  \mathcal{U}_i\mathcal{A}_i\;,
\label{eq:33}
\end{equation}
starting with $\mathcal{A}_1=\mathcal{I}$, the identity matrix. In
summary, the deflection-angle fields $\vec\alpha_i$ on the $N$ lens
planes can be used to construct the matrices $\mathcal{U}_i$ according
to Eq.~(\ref{eq:32}), then Eq.~(\ref{eq:33}) can be used to determine
the lensing experienced by a light ray starting out into any direction
on the image plane. The left panel in Fig.~\ref{eq:13} shows the total
convergence experienced by sources at $z_\mathrm{s}=5$ on a lens plane
with a side length of $4.25^\circ$, obtained from an $N$-body
simulation (Pfrommer 2002).

\begin{figure*}[ht]
  \includegraphics[width=0.49\hsize]{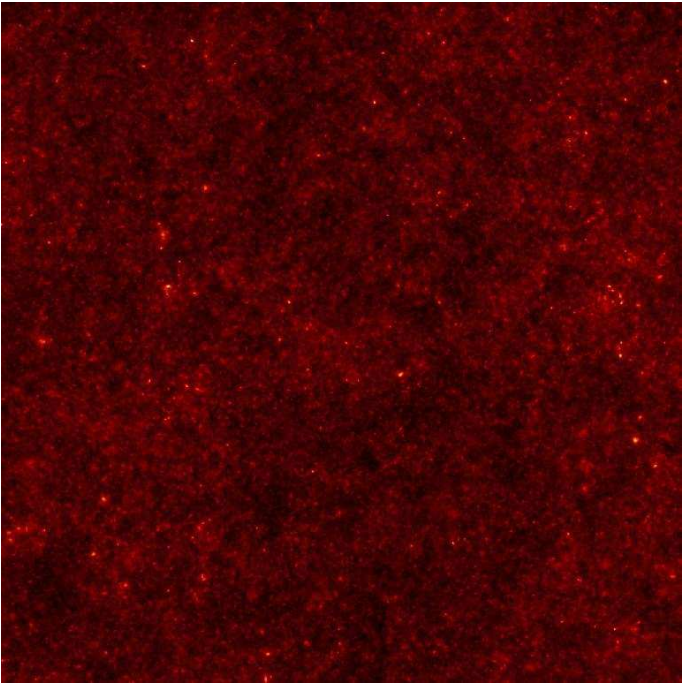}\hfill
  \includegraphics[width=0.49\hsize]{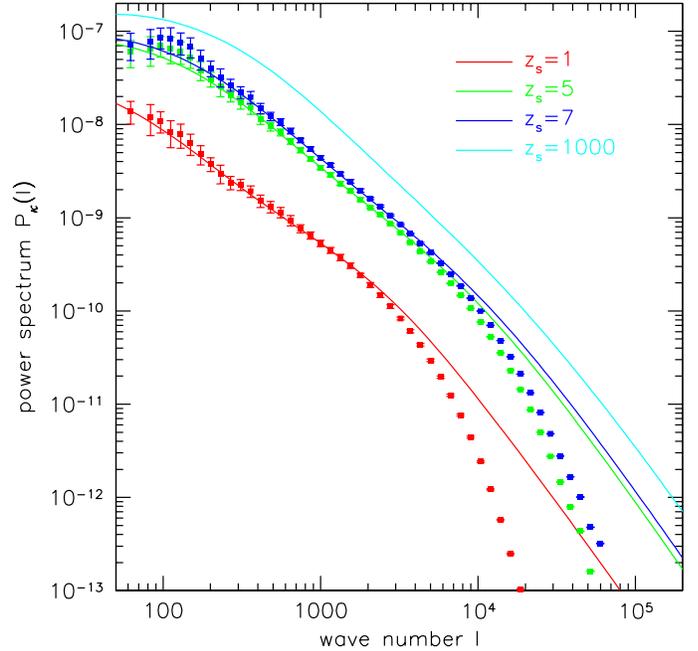}
\caption{\emph{Left panel}: Effective convergence on a field of
  $4.25^\circ$ side length for sources at redshift $z_\mathrm{s}=5$,
  obtained using multiple lens plane theory on an $N$-body
  simulation. \emph{Right panel}: Effective-convergence power spectra
  measured with the same set of simulations (crosses), compared with
  theoretical expectations (lines), for different source
  redshifts. The numerical results follow the theoretical curves very
  well within an intermediate range of wave numbers $l$. At larger
  $l$, i.e.~for small structures, the resolution limit of the
  simulation is reached and the power spectra fall rather steeply. At
  the low-$l$ end, the noise increases because the number of modes in
  the simulation decreases as the modes increase. The numerical power
  spectra for low-redshift sources fall below the theroretical
  expectation at lower $l$ than for high-redshift sources because a
  given angle, and thus wave number $l$, corresponds to a larger
  physical scale at smaller distances from the observer (from Pfrommer
  2002).}
\label{fig:13}
\end{figure*}

The right panel in Fig.~\ref{fig:13} shows numerically determined
power spectra for the effective convergence as functions of wave
number $l$, which is the Fourier conjugate variable to the angular
scale. The lines in this figure show the theoretically expected power
spectra. The agreement between the numerical and theoretical results
is very good over a limited range of wave numbers. Once the wave
numbers increase beyond the limit set by the angular resolution, the
simulated convergence fields lack power and the numerical results fall
below the theoretical ones. This happens at lower $l$ for smaller
source redshifts, because a fixed angular scale, and thus wave number
$l$, corresponds to smaller physical scales at lower distances. On the
low-$l$ end, i.e.~for large structures, the errors on the numerically
determined power spectra increase because the number of independent
modes in the simulated convergence field decreases as the modes
increase. This example should suffice to demonstrate that numerical
simulations of gravitational lensing by large-scale structures should
be carefully designed to match their final purpose.

\section{Inversion Techniques}

Let us conclude with a brief discussion of inversion techniques. They
are typically less demanding numerically, but the methods which have
been developed for this purpose are interesting in their own right.

\subsection{Shear Deconvolution}

We have seen before in Eqs.~(\ref{eq:23}) and (\ref{eq:24}) that
convergence and shear are related because they are both linear
combinations of second derivatives of the scalar lensing potential
$\psi$. In Fourier space, the relations are algebraic and can easily
be combined to eliminate the Fourier transform $\hat\psi$ of the
potential. Transforming back into configuration space, the convergence
turns out to be a convolution of the shear with a well-known kernel,
\begin{equation}
  \kappa(\vec\theta)=\frac{1}{\pi}\,
  \int\d^2\theta'\left[
    \mathcal{D}_1(\vec\theta-\vec\theta')\gamma_1(\vec\theta')+
    \mathcal{D}_2(\vec\theta-\vec\theta')\gamma_2(\vec\theta')
  \right]\;,
\label{eq:34}
\end{equation}
with
\begin{equation}
  \mathcal{D}_1(\vec\theta)=
  -\frac{\theta_1^2-\theta_2^2}{|\vec\theta|^4}\;,\quad
  \mathcal{D}_2(\vec\theta)=
  -\frac{2\theta_1\theta_2}{|\vec\theta|^4}\;.
\label{eq:35}
\end{equation}
This is the classic Kaiser \& Squires (1993) shear inversion
equation. Its limitations have been discussed in detail and removed to
satisfaction by modifying e.g.~the kernel components
$\mathcal{D}_{ij}$; they are not of interest for the discussion here
(cf.~Peter Schneider's lecture in this volume).

A suitable practical approximation of (\ref{eq:34}) using measured
galaxy ellipticities $\epsilon_i$ ($i=1,2$) is
\begin{equation}
  \kappa(\theta)\approx\frac{1}{n\pi}\,\sum_{i=1}^N\left[
    \mathcal{D}_1\epsilon_{1,i}+\mathcal{D}_2\epsilon_{2,i}
  \right]\;,
\label{eq:36}
\end{equation}
where $n$ is the number density of lensed galaxies on the sky. In
practice, however, it turns out that an approximation like
(\ref{eq:36}) would have infinite noise because of the random sampling
of the shear components $\gamma_i$ by $N$ galaxy ellipticities
$\epsilon_i$. This can be remedied by introducing a smoothed kernel
$\mathcal{D}'$ instead of $\mathcal{D}$, e.g.
\begin{equation}
  \mathcal{D}'=\left[
    1-\left(1+\frac{\theta^2}{\theta_\mathrm{s}^2}\right)
    \exp\left(-\frac{\theta^2}{\theta_\mathrm{s}^2}\right)
  \right]\,\mathcal{D}\;,
\label{eq:37}
\end{equation}
where $\theta_\mathrm{s}$ is the angular smoothing scale (Seitz \&
Schneider 1995). The noise convariance matrix between the convergence
values at two different grid points $\vec\theta_i$ and $\vec\theta_j$
is then
\begin{equation}
  \left\langle\kappa(\vec\theta_i)\kappa(\vec\theta_j)\right\rangle=
  \frac{\sigma_\epsilon^2}{4\pi\theta_\mathrm{s}^2n}\,
  \exp\left[
    -\frac{(\vec\theta_i-\vec\theta_j)^2}{2\theta_\mathrm{s}^2}
  \right]\;,
\label{eq:38}
\end{equation}
where $\sigma_\epsilon$ is the scatter of the intrinsic galaxy
ellipticities (van Waerbeke 2000). This expression demonstrates that
smoothing introduces correlations on the convergence map on the
angular scale $\theta_\mathrm{s}$, but the variance of $\kappa$ can
become very high if $\theta_\mathrm{s}$ is chosen too small. A careful
balance between the local variance and non-local correlations is
necessary in order to arrive at a convergence map with the required
properties.

\subsection{Maximum-Likelihood Lens Inversion}

An entirely different approach to lens inversion uses the
maximum-likelihood technique (Bartelmann et al.~1996). Each lensed
background galaxy $i$ provides a measurement of two ellipticity
components $(\epsilon_{1,i},\epsilon_{2,i})$ and its angular
size. Comparing the size of a galaxy behind a galaxy cluster to the
average size of unlensed galaxies of the same surface brightness, an
estimate $r_i$ of the inverse magnification of the lensed galaxy can
be derived. Thus $N$ galaxies provide a $3N$-dimensional data vector
\begin{equation}
  \vec d=(\epsilon_{1,1},\epsilon_{2,1},r_1,\ldots,
          \epsilon_{1,N},\epsilon_{2,N},r_N)\;.
\label{eq:39}
\end{equation}
The goal of the lens inversion is then to find a two-dimensional array
$\psi_{jk}$ of lensing potential values such that the ellipticities
and inverse magnifications caused by that potential at the positions
$\vec\theta_i$ of the real galaxies optimally reproduce the measured
ellipticities and inverse magnifications. In other words, the
potential values $\psi_{jk}$ have to be determined such as to minimise
the mean-square difference between the data vector $\vec d$ and the
model data vector $\vec d[\psi_{jk}(\vec x_i)]$,
\begin{equation}
  \chi^2(\psi_{jk})=\sum_{i=1}^{3N}\left\{
    \frac{[d_i-d_i(\psi_{jk})]^2}{\sigma_i^2}
  \right\}\;,
\label{eq:40}
\end{equation}
where the errors $\sigma_i$ can be estimated from the data
themselves. The minimisation of $\chi^2$ with respect to the potential
values $\psi_{jk}$ can be done with any minimisation algorithm like,
e.g.~the downhill simplex method. For large fields, the number of
potential values can become very large. In that case,
conjugate-gradient methods are preferred, which make use of the fact
that the derivatives of $\chi^2$ with respect to the $\psi_{jk}$ are
known analytically. Such methods can speed up the minimisation
sufficiently to render it feasible even for large potential arrays
(cf.~Press et al.~1992).

\subsection{Maximum-Entropy Methods}

The minimisation of $\chi^2$ is a special case of the
maximum-likelihood technique for assumed Gaussian deviations of the
measured data around the model values. Improvements of the
maximum-likelihood technique can be derived starting from Bayes'
theorem,
\begin{equation}
  P(\psi|\vec d)=\frac{P(\vec d|\psi)}{P(\vec d)}\,P(\psi)\;,
\label{eq:41}
\end{equation}
which states that the probability $P(\psi|\vec d)$ of finding the
potential $\psi$ given the data $\vec d$ is proportional to the
probability $P(\vec d|\psi)$ of obtaining the data given the
potential, times the probability $P(\psi)$ for finding the
potential. The denominator $P(\vec d)$ is called the \emph{evidence}
and simply normalises Eq.~(\ref{eq:41}). $P(\psi)$ is called the
prior, quantifying any \emph{a priori} information one has or assumes
on the potential $\psi$, $P(\vec d|\psi)$ is called the likelihood,
and $P(\psi|\vec d)$ is the \emph{posterior} probability. The goal is
now to maximise the latter, which is equivalent to maximising the
product $P(\vec d|\psi)\,P(\psi)$ of likelihood and prior. If we have
or can assume Gaussian noise and a diagonal noise correlation matrix,
the likelihood reduces to $P(\vec d|\psi)=\exp(-\chi^2/2)$.

It can now be shown that in absence of any further information, the
best, i.e.~least prejudiced, prior is the maximum-entropy prior,
\begin{equation}
  P(\psi)\propto\exp\left[\alpha\,S(\psi,\vec m)\right]\;,
\label{eq:42}
\end{equation}
with the \emph{cross entropy}
\begin{equation}
  S(\psi,\vec m)=\sum_{i=1}^{3N}\,
  \psi_i-m_i-\psi_i\,\ln\frac{\psi_i}{m_i}\;,
\label{eq:43}
\end{equation}
where $\vec m$ is a model vector for the potential which can encode
expectations on the potential, or simply be chosen to be uniform for
all $i$. The potential array is then determined by maximising
$\exp(-\chi^2/2+\alpha S)$, or equivalently by minimising
\begin{equation}
  F\equiv\frac{1}{2}\chi^2-\alpha S
\label{eq:44}
\end{equation}
instead of the simple $\chi^2$ in Eq.~(\ref{eq:40}). The parameter
$\alpha$ can be included into the minimisation. Bayesian theory
implies that a good approximation to the optimal choice for $\alpha$
is determined such that $F\sim3N/2$ at the potential minimum
$\bar\psi$. The error covariance matrix for the potential $\psi$ is
given by the inverse curvature matrix of $F$,
\begin{equation}
  \left\langle(\psi-\bar\psi)(\psi-\bar\psi)^\mathrm{T}\right\rangle
  \approx
  \left(\frac{\partial^2F}{\partial\psi_i\partial\psi_j}\right)^{-1}\;.
\label{eq:45}
\end{equation}
Maximum-entropy methods have been suggested and used for regularising
shear-inversion techniques such that their spatial resolution is
adapted to the strength of the lensing signal (Bridle et al.~1998;
Seitz et al.~1998).

\section{Concluding Remarks}

Many numerical methods have been used for gravitational lensing
studies which I was not able to cover during the limited time of the
lecture. Among them are the hierarchical tree-code methods introduced
into microlensing by Wambsganss et al.~(1990) and the methods for
constraining cluster mass distributions from multiple arc systems
(e.g.~Kneib et al.~1993; see also Jean-Paul Kneib's presentation in
this volume). Despite this unavoidable incompleteness, I hope to have
given a flavour of how numerical methods can be used for lensing, and
what the main problem areas are.

\acknowledgements{I wish to thank the organisers of the lensing school
in Aussois which provided a nice opportunity to discuss a broad
variety of lensing topics, and a scenic frame.}


\begin{thebibliography}{99}

\bibitem{} Bartelmann, M., Weiss, A. 1994, A\&A 287, 1
\bibitem{} Bartelmann, M., Steinmetz, M., Weiss, A. 1995, A\&A 297, 1
\bibitem{} Bartelmann, M., Narayan, R., Seitz, S., Schneider, P. 1996,
  ApJ 464, L115
\bibitem{} Bartelmann, M., Loeb, A. 1998, ApJ 503, 48
\bibitem{} Bartelmann, M., Schneider, P. 2001, Phys. Rep. 340, 291
\bibitem{} Bridle, S.L., Hobson, M.P., Lasenby, A.N., Saunders,
  R. 1998, MNRAS 299, 895
\bibitem{} Hockney, R.W., Eastwood, J.W. 1988, Computer Simulations
  using Particles (Bristol: Hilger)
\bibitem{} Kaiser, N., Squires, G. 1993, ApJ 404, 441
\bibitem{} Keeton, C.R., Kochanek, C.S. 1998, ApJ 495, 157
\bibitem{} Kneib, J.-P., Mellier, Y., Fort, B., Mathez, G. 1993, A\&A
  273, 367
\bibitem{} Mellier, Y. 1999, Ann. Rev. Astr. Ap. 37, 127
\bibitem{} Meneghetti, M., Yoshida, N., Bartelmann, M., Moscardini,
  L. Springel, V., Tormen, G., White, S.D.M. 2001, MNRAS 325, 435
\bibitem{} Meneghetti, M., Bolzonella, M., Bartelmann, M., Moscardini,
  L., Tormen, G. 2000, MNRAS 314, 338
\bibitem{} Narayan, R., Bartelmann, M. 1999, in: Formation of
  Structure in the Universe, eds. A. Dekel and J.P. Ostriker, p. 360
  (Cambridge: University Press)
\bibitem{} Pfrommer, C. 2002, \emph{Diploma Thesis}, Munich University
\bibitem{} Press, W.H., Teukolsky, S.A., Vetterling, W.T., Flannery,
  B.P. 1992, Numerical Recipes (Cambridge: University Press)
\bibitem{} Schneider, P., Ehlers, J., Falco, E.E. 1992, Gravitational
  Lenses (Heidelberg: Springer Verlag)
\bibitem{} Seitz, C., Schneider, P. 1995, A\&A 297, 287
\bibitem{} Seitz, S., Schneider, P., Bartelmann, M. 1998, A\&A 337,
  325
\bibitem{} Seljak, U., Zaldarriaga, M. 2000, ApJ 538, 57
\bibitem{} van Waerbeke, L. 2000, MNRAS 313, 524
\bibitem{} Wambsganss, J., Paczy\'nski, B., Katz, N. 1990, ApJ 352,
  407

\end{thebibliography}
\end{document}